\pdfoutput=1
\documentclass[prd,showpacs,letterpaper,10pt,aps,notitlepage,superscriptaddress,nofootinbib,preprintnumbers]{revtex4-1}

\usepackage{amsfonts,amsmath,graphicx}
\usepackage{slashed}
\usepackage{placeins}

\newcommand{\nb}{\phantom{0}}
\newcommand{\wm}{\phantom{-}}
\newcommand{\bs}[1]{\ensuremath{{\boldsymbol{#1}}}}

\begin{document}

\title{$\Lambda_b \to p\: \ell^-\, \bar{\nu}_\ell$ form factors from lattice QCD with static $b$ quarks}

\author{William Detmold}
\affiliation{Center for Theoretical Physics, Massachusetts Institute of Technology, Cambridge, MA 02139, USA}
\author{C.-J.~David Lin}
\affiliation{Institute of Physics, National Chiao-Tung University, Hsinchu 300, Taiwan}
\author{Stefan Meinel}
\email{smeinel@mit.edu}
\affiliation{Center for Theoretical Physics, Massachusetts Institute of Technology, Cambridge, MA 02139, USA}
\author{Matthew Wingate}
\affiliation{DAMTP, University of Cambridge, Wilberforce Road, Cambridge CB3 0WA, UK}


\date{June 3, 2013}

\preprint{MIT-CTP 4465}

\begin{abstract}
We present a lattice QCD calculation of form factors for the decay $\Lambda_b \to p\: \mu^- \bar{\nu}_\mu$, which
is a promising channel for determining the CKM matrix element $|V_{ub}|$ at the Large Hadron Collider.
In this initial study we work in the limit of static $b$ quarks, where the number of independent form factors reduces to two.
We use dynamical domain-wall fermions for the light quarks, and perform the calculation at two different lattice spacings and
at multiple values of the light-quark masses in a single large volume. Using our form factor results, we calculate
the $\Lambda_b \to p\: \mu^- \bar{\nu}_\mu$ differential decay rate in the range $14\hspace{1ex}{\rm GeV}^2 \leq q^2 \leq q^2_{\rm max}$,
and obtain the integral
$\int_{14\:{\rm GeV}^2}^{q^2_{\rm max}} [\mathrm{d}\Gamma/\mathrm{d} q^2] \mathrm{d} q^2 / |V_{ub}|^2 = 15.3 \pm 4.2 \:\: \rm{ps}^{-1}$.
Combined with future experimental data, this will give a novel determination of $|V_{ub}|$ with about 15\% theoretical uncertainty. The
uncertainty is dominated by the use of the static approximation for the $b$ quark, and can be reduced further
by performing the lattice calculation with a more sophisticated heavy-quark action.
\end{abstract}

\maketitle

\FloatBarrier
\section{Introduction}
\FloatBarrier

A long-standing puzzle in flavor physics is the discrepancy between the extractions of the CKM matrix element $|V_{ub}|$
from inclusive and exclusive $B$ meson semileptonic decays at the $B$ factories \cite{Kowalewski:2010zz, Mannel:2010zz,
Antonelli:2009ws, PDG2012}. The current average values determined by the Particle Data Group are \cite{PDG2012}
\begin{eqnarray}
 |V_{ub}|_{\rm incl.} &=& (4.41 \pm 0.15 ^{+0.15}_{-0.17} )\cdot 10^{-3}, \label{eq:Vubincl} \\
 |V_{ub}|_{\rm excl.} &=& (3.23 \pm 0.31 )\cdot 10^{-3}, \label{eq:Vubexcl}
\end{eqnarray}
where the exclusive determination is based on measurements of $\bar{B} \to \pi^+ \ell^- \bar{\nu}_\ell$ decays by the BABAR
and BELLE collaborations, and uses $\bar{B} \to \pi^+$ form factors computed in lattice QCD \cite{Bailey:2008wp, Dalgic:2006dt}.
To address the discrepancy between Eqs.~(\ref{eq:Vubincl}) and (\ref{eq:Vubexcl}), new and independent determinations of
$|V_{ub}|$ are desirable. At the Large Hadron Collider, measurements of $\bar{B} \to \pi^+ \ell^- \bar{\nu}_\ell$ branching fractions
are difficult because of the large pion background; therefore an attractive possibility is to use instead the baryonic mode
$\Lambda_b \to p\: \ell^- \bar{\nu}_\ell$, which has a more distinctive final state \cite{Egede}. In order to
determine $|V_{ub}|$ from this measurement, the $\Lambda_b \to p$ form factors need to be calculated in nonperturbative QCD.

The $\Lambda_b \to p$ matrix elements of the vector and axial vector $b\to u$ currents are parametrized in terms of six
independent form factors (see, e.g., Ref.~\cite{Hussain:1992rb}). In leading-order heavy-quark effective theory (HQET),
which becomes exact in the limit $m_b\to \infty$ and is a good approximation at the physical value of $m_b$,
only two independent form factors remain, and the matrix element with arbitrary Dirac matrix $\Gamma$ in the current
can be written as \cite{Mannel:1990vg, Hussain:1990uu, Hussain:1992rb}
\begin{equation}
\langle N^+(p', s') | \:\bar{u} \Gamma Q \: | \Lambda_Q(v, s) \rangle =
\overline{u}_N(p',s')\left[ F_1 + \slashed{v}\:F_2 \right] \Gamma\: u_{\Lambda_Q}(v, s). \label{eq:FFdef}
\end{equation}
Above, $v$ is the four-velocity of the $\Lambda_Q$ baryon, and the form factors $F_1$, $F_2$ are functions of $p'\cdot v$,
the energy of the proton in the $\Lambda_Q$ rest frame (we denote the heavy quark defined in HQET by $Q$, and we denote
the proton by $N^+$). Note that in leading-order soft-collinear effective theory, which applies in the limit of large $p'\cdot v$,
the form factor $F_2$ vanishes \cite{Feldmann:2011xf, Mannel:2011xg, Wang:2011uv}.

Calculations of $\Lambda_b \to p$ form factors have been performed using QCD sum rules \cite{Huang:1998rq, Carvalho:1999ia}
and light-cone sum rules \cite{Huang:2004vf, Wang:2009hra, Azizi:2009wn, Khodjamirian:2011jp}. Light-cone sum rules
are most reliable at low $q^2$ (corresponding to large proton momentum in the $\Lambda_b$ rest frame),
and even there the uncertainty of the best available calculations is of order 20\% \cite{Khodjamirian:2011jp}.
As we will see later, the $\Lambda_b \to p\: \ell^- \bar{\nu}_\ell$ differential decay rate
has its largest value in the high-$q^2$ (low hadronic recoil) region. This is also the region
where lattice QCD calculations can be performed with the highest precision.

Lattice QCD determinations of the form factors for the mesonic decay $\bar{B} \to \pi^+ \ell^- \bar{\nu}_\ell$
are already available \cite{Dalgic:2006dt, Bailey:2008wp}, and several groups are working on new calculations
\cite{Bahr:2012vt, Bouchard:2012tb, Kawanai:2012id}. We have recently published the first lattice QCD calculation of
$\Lambda_Q \to \Lambda$ form factors, which are important for the rare decay $\Lambda_b \to \Lambda\: \ell^+\ell^-$
\cite{Detmold:2012vy}. We performed this calculation at leading order in HQET, i.e., with static heavy quarks. The HQET
form factors $F_1$ and $F_2$ for the $\Lambda_Q \to \Lambda$ transition are defined as in Eq.~(\ref{eq:FFdef}), except
that the current is $\bar{s}\Gamma Q$ and the final state is the $\Lambda$ baryon.
In the following, we report the first lattice QCD determination of the $\Lambda_Q \to p$ form factors defined in
Eq.~(\ref{eq:FFdef}), building upon the analysis techniques developed in Ref.~\cite{Detmold:2012vy}. The calculation
uses dynamical domain wall fermions \cite{Kaplan:1992bt, Furman:1994ky, Shamir:1993zy} for the up-, down-, and strange
quarks, and is based on gauge field ensembles generated by the RBC/UKQCD collaboration \cite{Aoki:2010dy}.

In Sec.~\ref{sec:latticecalc}, we outline our extraction of the $\Lambda_Q \to p$ form factors from ratios of correlation
functions, and present the lattice parameters and form factor results for each data set.
In Sec.~\ref{sec:chiralcontinuumextrap}, we present our fits of the lattice-spacing-, quark-mass-, and $(p'\cdot v)$-dependence
of these results, and discuss systematic uncertainties. We compare the $\Lambda_Q \to p$ form factors
computed here to previous determinations of both $\Lambda_Q \to p$ and $\Lambda_Q \to \Lambda$ form factors in
Sec.~\ref{sec:FFcomparison}. Using our form factor results, we then calculate the differential decay rates of
$\Lambda_b \to p\: \ell^- \bar{\nu}_\ell$ for $\ell=e,\mu,\tau$ in Sec.~\ref{sec:Lambdabdecay}.
Finally, in Sec.~\ref{sec:conclusions} we discuss the impact of our results on future determinations of $|V_{ub}|$
from these decays, and the prospects for more precise lattice calculations.

\FloatBarrier
\section{\label{sec:latticecalc}Lattice calculation}
\FloatBarrier

We performed the calculation of the $\Lambda_Q \to p$ form factors with the same lattice actions and parameters as
used in our calculation of the $\Lambda_Q \to \Lambda$ form factors in Ref.~\cite{Detmold:2012vy}. That is,
we are using a domain-wall action for the up, down, and strange quarks  \cite{Kaplan:1992bt, Furman:1994ky, Shamir:1993zy},
the Iwasaki action \cite{Iwasaki:1983ck, Iwasaki:1984cj} for the gluons, and the Eichten-Hill action \cite{Eichten:1989kb}
with HYP-smeared gauge links \cite{DellaMorte:2005yc} for the static heavy quark. The Eichten-Hill action
requires that we work in the $\Lambda_Q$ rest frame, i.e.~with $v=(1,0,0,0)$. We compute ``forward'' and ``backward''
three-point functions
\begin{eqnarray}
 C^{(3)}_{\delta\alpha}(\Gamma,\:\mathbf{p'}, t, t') &=& \sum_{\mathbf{y}} e^{-i\mathbf{p'}\cdot(\mathbf{x}-\mathbf{y})}
 \left\langle N_{\delta}(x_0,\mathbf{x})\:\: J_\Gamma^{\dag}(x_0-t+t',\mathbf{y})
 \:\:\: \overline{\Lambda}_{Q\alpha} (x_0-t,\mathbf{y}) \right\rangle, \label{eq:threept} \\
C^{(3,\mathrm{bw})}_{\alpha\delta}(\Gamma,\:\mathbf{p'}, t, t-t') &=& \sum_{\mathbf{y}}
e^{-i\mathbf{p'}\cdot(\mathbf{y}-\mathbf{x})} \left\langle \Lambda_{Q\alpha}(x_0+t,\mathbf{y})\:\: J_\Gamma(x_0+t',\mathbf{y})
\:\:\: \overline{N}_{\delta} (x_0,\mathbf{x}) \right\rangle, \label{eq:threeptbw}
\end{eqnarray}
containing the baryon interpolating fields
\begin{eqnarray}
 \Lambda_{Q\alpha} &=& \epsilon^{abc}\:(C\gamma_5)_{\beta\gamma}\:\tilde{d}^a_\beta\:\tilde{u}^b_\gamma\: Q^c_\alpha, \label{eq:lambdaqinterpol} \\
 N_{\alpha} &=& \epsilon^{abc}\:(C\gamma_5)_{\beta\gamma}\:\tilde{u}^a_\beta\:\tilde{d}^b_\gamma\: \tilde{u}^c_\alpha, \label{eq:protoninterpol}
\end{eqnarray}
and the current
\begin{equation}
 J_{\Gamma} = U(m_b, a^{-1})\: \mathcal{Z}
 \left[ \left( 1  + c^{(m a)}_\Gamma\:\frac{m_u\:a}{1-(w_0^{\rm MF})^2}\right) \overline{Q} \Gamma  u
 + c^{(p a)}_\Gamma\:a\: \overline{Q} \Gamma \bs{\gamma}\cdot \bs{\nabla}  u   \right]. \label{eq:LHQETcurrent}
\end{equation}
In the baryon interpolating fields, the tilde on the up- and down-quark fields indicates Gaussian gauge-covariant smearing.
The coefficients $\mathcal{Z}$, $c^{(m a)}_\Gamma$, and $c^{(p a)}_\Gamma$ in the current (\ref{eq:LHQETcurrent}) provide an
$\mathcal{O}(a)$-improved matching from lattice HQET to continuum HQET in the $\overline{\rm MS}$ scheme; they have been computed
in one-loop perturbation theory in Ref.~\cite{Ishikawa:2011dd}. The factor $U(m_b, a^{-1})$ provides two-loop renormalization-group
running in continuum HQET from the scale $\mu=a^{-1}$ (where $a$ is the lattice spacing) to the desired scale $\mu=m_b$.

Note that, because Eq.~(\ref{eq:protoninterpol}) contains two up-quark fields, the $\Lambda_Q \to p$ three-point functions
contain two different types of contractions of quark propagators, one of which is not present in the $\Lambda_Q \to \Lambda$
three-point functions studied in Ref.~\cite{Detmold:2012vy}. This is also the case for the proton two-point functions.

We multiply the forward- and backward three-point functions and form the ratio \cite{Detmold:2012vy}
\begin{equation}
 \mathcal{R}(\Gamma, \mathbf{p'}, t, t') = \frac{4 \:\mathrm{Tr}\left[  C^{(3)}(\Gamma,\:\mathbf{p'}, t, t')
 \:\: C^{(3,\mathrm{bw})}(\Gamma,\:\mathbf{p'}, t, t-t') \right] }{ \mathrm{Tr}[ C^{(2,N)}(\mathbf{p'}, t)]
 \:\mathrm{Tr}[ C^{(2,\Lambda_Q)}(t) ] }, \label{eq:doubleratio}
\end{equation}
where $C^{(2,N)}(\mathbf{p'}, t)$ and $C^{(2,\Lambda_Q)}(t)$ are the proton and the $\Lambda_Q$ two-point functions, and
the traces are over spinor indices. The ratio is computed using the statistical bootstrap method. As explained in
Ref.~\cite{Detmold:2012vy}, we then form the combinations
\begin{eqnarray}
 \mathcal{R}_+(\mathbf{p'}, t, t') &=& \frac14 \left[ \mathcal{R}(1, \mathbf{p'}, t, t') + \mathcal{R}(\gamma^2\gamma^3, \mathbf{p'}, t, t')
 + \mathcal{R}(\gamma^3\gamma^1, \mathbf{p'}, t, t') + \mathcal{R}(\gamma^1\gamma^2, \mathbf{p'}, t, t') \right],  \label{eq:curlyRplus} \\
 \mathcal{R}_-(\mathbf{p'}, t, t') &=& \frac14 \left[ \mathcal{R}(\gamma^1, \mathbf{p'}, t, t') + \mathcal{R}(\gamma^2, \mathbf{p'}, t, t')
 + \mathcal{R}(\gamma^3, \mathbf{p'}, t, t') + \mathcal{R}(\gamma_5, \mathbf{p'}, t, t') \right],  \label{eq:curlyRminus}
\end{eqnarray}
which, upon inserting Eq.~(\ref{eq:FFdef}) into the transfer matrix formalism, yield
\begin{eqnarray}
 \mathcal{R}_+(\mathbf{p'}, t, t') &=& \frac{ E_N+m_N}{E_N} [F_1 + F_2]^2 + \hdots, \label{eq:RGammap}\\
 \mathcal{R}_-(\mathbf{p'}, t, t') &=& \frac{ E_N-m_N}{E_N} [F_1 - F_2]^2 + \hdots. \label{eq:RGammam}
\end{eqnarray}
Here, the ellipses denote excited-state contributions that decay exponentially with the Euclidean time separations, and
$F_1$, $F_2$ are the form factors at the given values of the proton momentum $\mathbf{p'}$, the lattice spacing, and the
quark masses. Throughout the remainder of this paper, we will use the following names for the combinations of form factors
that appear in Eqs.~(\ref{eq:RGammap}), (\ref{eq:RGammam}):
\begin{equation}
F_+ = F_1 + F_2, \hspace{4ex} F_- = F_1 - F_2.
\end{equation}
For a given value of $|\mathbf{p'}|^2$, we further average Eqs.~(\ref{eq:curlyRplus}) and (\ref{eq:curlyRminus}) over the
direction of $\mathbf{p'}$, and we denote the resulting quantities as $\mathcal{R}_\pm (|\mathbf{p'}|^2, t, t')$.
As a consequence of the symmetric form of the ratio (\ref{eq:doubleratio}), at a given source-sink separation $t$,
the contamination from excited states is smallest at the mid-point $t'=t/2$. We therefore construct the following functions
\begin{eqnarray}
 R_+(|\mathbf{p'}|^2, t) &=& \sqrt{\frac{E_N}{E_N+m_N} \mathcal{R}_+(|\mathbf{p'}|^2,\: t,\: t/2)}, \label{eq:Rplus}\\
 R_-(|\mathbf{p'}|^2, t) &=& \sqrt{\frac{E_N}{E_N-m_N} \mathcal{R}_-(|\mathbf{p'}|^2,\: t,\: t/2)}, \label{eq:Rminus}
\end{eqnarray}
which, according to Eqs.~(\ref{eq:RGammap}), (\ref{eq:RGammam}), become equal to the form factors $F_+$ and $F_-$
for large source-sink separation, $t$.

We performed the numerical calculations for the six different sets of parameters shown in Table \ref{tab:params}.
When evaluating Eqs.~(\ref{eq:Rplus}) and (\ref{eq:Rminus}), we used the lattice results for the proton mass, $m_N$, obtained
from fits to the proton two-point function in the same data set. These results are also given in Table \ref{tab:params}.
Unlike in Ref.~\cite{Detmold:2012vy}, here we calculated the energies at nonzero momentum using the relativistic continuum
dispersion relation $E_N=\sqrt{m_N^2 + |\mathbf{p'}|^2}$. The energies calculated in this way are consistent with the
energies obtained directly from fits to the proton two-point functions at nonzero momentum, but using the relativistic
dispersion relation reduces the uncertainty. We computed $R_\pm(|\mathbf{p'}|^2, t)$ for proton momenta in the range
$0 \leq |\mathbf{p'}|^2 \leq 9\cdot (2\pi/L)^2$, where $L=N_s a \approx 2.7$ fm is the spatial size of the lattice.
We performed the calculation for all source-sink separations from $t/a=4$ to $t/a=15$ at the coarse lattice spacing
(data sets \texttt{C14}, \texttt{C24}, \texttt{C54}), and for $t/a=5$ to $t/a=20$ at the fine lattice spacing
(data sets \texttt{F23}, \texttt{F43}, \texttt{F63}). This wide range of source-sink separations allows us to reliably
extract the ground-state form factors \cite{Detmold:2012vy}. Because the statistical uncertainties grow exponentially
with $t$, in practice the upper limit of $t/a$ we can use is somewhat smaller, especially at larger momentum.

\begin{figure}
\vspace{10ex}
 \includegraphics[height=6cm]{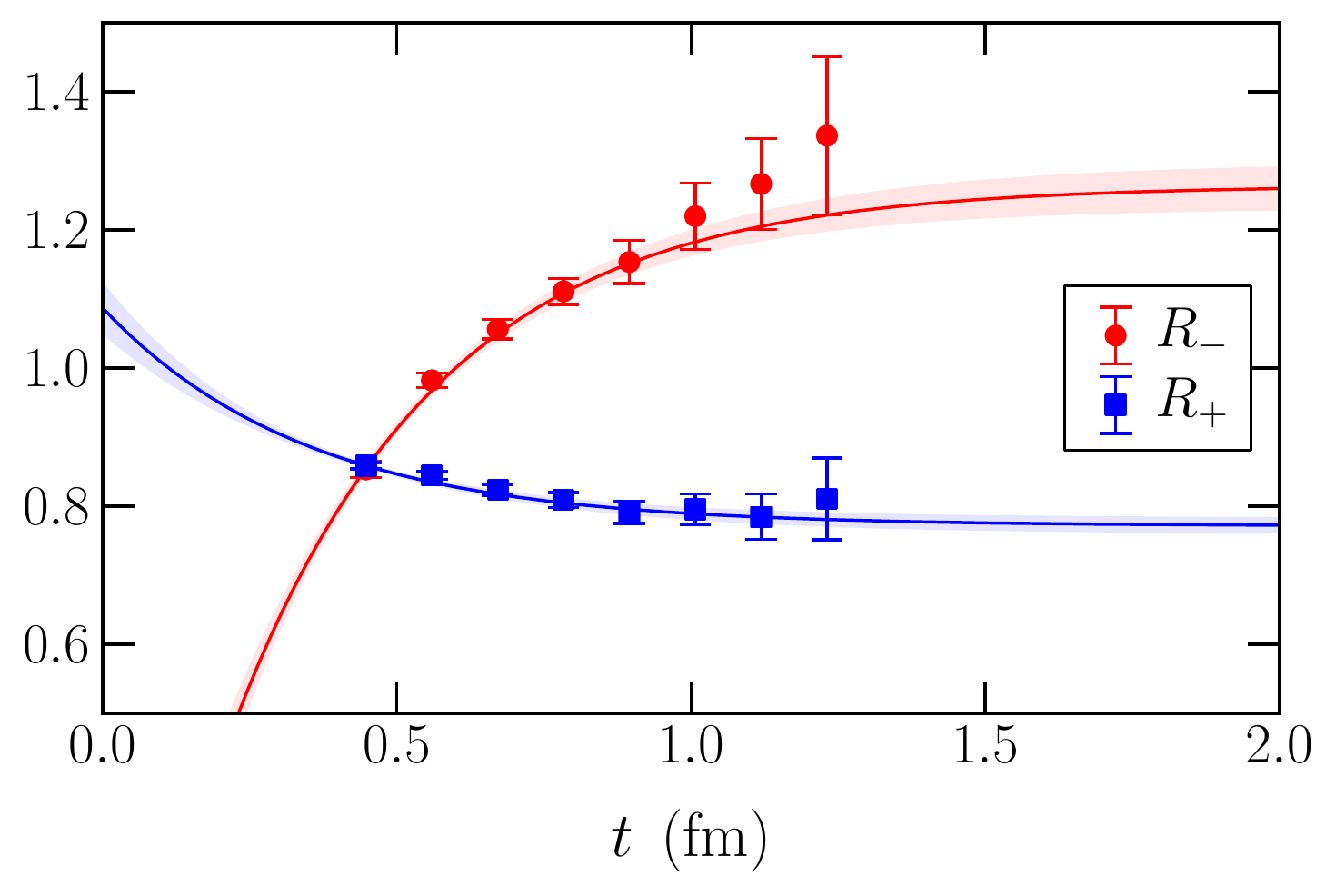}
 \caption{\label{fig:textrap_L24_005_psqr4}Example of numerical results for $R_\pm(|\mathbf{p'}|^2, t)$, plotted as
 a function of the source-sink separation $t$, along with a fit using Eq.~(\ref{eq:tdepsepcoupled}).
The data shown here are from the \texttt{C54} set and at $|\mathbf{p'}|^2=4\cdot(2\pi/L)^2$. As explained in
Ref.~\protect\cite{Detmold:2012vy}, at each value of $|\mathbf{p'}|^2$, the fit is performed simultaneously
for the six data sets.}
\vspace{10ex}
\end{figure}

A plot of example numerical results for $R_\pm(|\mathbf{p'}|^2, t)$ as a function of the source-sink separation $t$ is
shown in Fig.~\ref{fig:textrap_L24_005_psqr4}. The results are qualitatively similar to those obtained for the $\Lambda$
final state in Ref.~\cite{Detmold:2012vy} (the $t'$-dependence of $\mathcal{R}_\pm (|\mathbf{p'}|^2, t, t')$ is also
similar to that seen in Ref.~\cite{Detmold:2012vy}). It can be seen that there is contamination from excited states
which decays exponentially with $t$. At $|\mathbf{p'}|^2\neq0$ we perform fits of the $t$-dependence using the functions
\begin{equation}
R^{i,n}_\pm(t) = F^{i,n}_\pm + A^{i,n}_\pm \:\exp[-\delta^{i,n}\:t], \label{eq:tdepsepcoupled}
\end{equation}
which account for the leading excited-state contamination \cite{Detmold:2012vy}. Above, we use an abbreviated notation
where $n$ specifies the squared momentum of the proton [we write $|\mathbf{p'}|^2 = n\cdot (2\pi/L)^2$], and
$i=\mathtt{C14}, \mathtt{C24}, ..., \mathtt{F63}$ specifies the data set. To enforce the positivity of the energy gaps
$\delta^{i,n}$, we rewrite them as $\delta^{i,n} /(1\: {\rm GeV}) = \exp(l^{i,n})$. The fit parameters in
Eq.~(\ref{eq:tdepsepcoupled}) are then $F^{i,n}_\pm$, $A^{i,n}_\pm$, and $l^{i,n}$. Note that we perform coupled fits of
$R^{i,n}_+$ and $R^{i,n}_-$ with common energy gap parameters, which improves the statistical precision of the fits
\cite{Detmold:2012vy}. As a check, we have also performed independent fits with separate energy gap parameters $l^{i,n}_+$
and $l^{i,n}_-$ and found that $l^{i,n}_+$ and $l^{i,n}_-$ are in agreement within statistical uncertainties.

\begin{table}
\begin{tabular}{cccccccccccccccccccccc}
\hline\hline
Set & \hspace{1ex} & $\beta$ & \hspace{1ex} & $N_s^3\times N_t\times N_5$ & \hspace{1ex} & $a m_5$ & \hspace{1ex} & $am_{s}^{(\mathrm{sea})}$
& \hspace{1ex} & $am_{u,d}^{(\mathrm{sea})}$   & \hspace{1ex} & $a$ (fm) & \hspace{1ex} & $am_{u,d}^{(\mathrm{val})}$ 
& \hspace{1ex} & $m_\pi^{(\mathrm{val})}$ (MeV) & \hspace{1ex} & $m_N^{(\mathrm{val})}$ (MeV) & \hspace{1ex} & $N_{\rm meas}$ \\
\hline
\texttt{C14} && $2.13$ && $24^3\times64\times16$ && $1.8$ && $0.04$ && $0.005$ && $0.1119(17)$ && $0.001$ && 245(4) && 1090(21) && 2672 \\
\texttt{C24} && $2.13$ && $24^3\times64\times16$ && $1.8$ && $0.04$ && $0.005$ && $0.1119(17)$ && $0.002$ && 270(4) && 1103(20) && 2676 \\
\texttt{C54} && $2.13$ && $24^3\times64\times16$ && $1.8$ && $0.04$ && $0.005$ && $0.1119(17)$ && $0.005$ && 336(5) && 1160(19) && 2782 \\
\texttt{F23} && $2.25$ && $32^3\times64\times16$ && $1.8$ && $0.03$ && $0.004$ && $0.0849(12)$ && $0.002$ && 227(3) && 1049(25) && 1907 \\
\texttt{F43} && $2.25$ && $32^3\times64\times16$ && $1.8$ && $0.03$ && $0.004$ && $0.0849(12)$ && $0.004$ && 295(4) && 1094(18) && 1917 \\
\texttt{F63} && $2.25$ && $32^3\times64\times16$ && $1.8$ && $0.03$ && $0.006$ && $0.0848(17)$ && $0.006$ && 352(7) && 1165(23) && 2782 \\
\hline\hline
\end{tabular}
\caption{\label{tab:params} Lattice parameters.
The data sets \texttt{C14}, \texttt{C24} and \texttt{C54}
all correspond to the same ``coarse'' ensemble of gauge fields with gauge coupling $\beta=6/g^2=2.13$ and sea-quark masses
$am_{s}^{(\mathrm{sea})}=0.04$, $am_{u,d}^{(\mathrm{sea})}=0.005$; these data sets differ only in the values of the
valence-quark mass, $am_{u,d}^{(\mathrm{val})}$, used for the domain wall propagators. At the ``fine'' lattice spacing,
the propagators in the \texttt{F23} and \texttt{F43} data sets are from one common ensemble of gauge fields, but the
\texttt{F63} data set is obtained from a different ensemble with heavier sea-quark masses. In each case, we also list
the valence pion and proton masses, $m_\pi^{(\mathrm{val})}$ and $m_N^{(\mathrm{val})}$, and the number of light-quark
propagators, $N_{\rm meas}$, used for our analysis. The ensembles of gauge fields have been generated by the RBC/UKQCD
collaboration; see Ref.~\cite{Aoki:2010dy} for further details.}
\end{table}

At a given momentum-squared $n$, we perform the fits using Eq.~(\ref{eq:tdepsepcoupled}) simultaneously for the six
different data sets $i=\mathtt{C14}, \mathtt{C24}, ..., \mathtt{F63}$. Because the lattice size, $L$ (in physical units),
is equal within uncertainties for all data sets, the squared momentum $|\mathbf{p'}|^2= n\cdot (2\pi/L)^2$ for a given
$n$ is also equal within uncertainties for all data sets. To improve the stability of the fits, we augment the $\chi^2$ function
by adding a term that limits the variation of $l^{i,n}$ across the data sets to reasonable values \cite{Detmold:2012vy}.

At $\mathbf{p'}=0$, we can only compute $R^{i,0}_+(t)$, and we find that the $t$-dependence of $R^{i,0}_+(t)$ is weak.
In this case we are unable to perform exponential fits, and we instead perform constant fits, excluding a few points
at the shortest $t$.

The numerical results for the form factors $F^{i,n}_\pm$ are listed in Tables \ref{tab:Fplus} and \ref{tab:Fminus}.
The uncertainties shown there are the quadratic combination of the statistical uncertainty and an estimate of the
systematic uncertainty associated with the choice of fit range for Eq.~(\ref{eq:tdepsepcoupled}). To estimate this
systematic uncertainty, we calculated the changes in the fitted $F^{i,n}_\pm$ when excluding the data points with
the shortest source-sink separation \cite{Detmold:2012vy}. These changes fluctuate as a function of the momentum,
and here we conservatively took the maximum of the change over all momenta as the systematic uncertainty for each
data set. In Tables \ref{tab:F1} and \ref{tab:F2}, we additionally list the corresponding results for
$F^{i,n}_1 = (F^{i,n}_+ + F^{i,n}_-)/2$ and $F^{i,n}_2=(F^{i,n}_+ - F^{i,n}_-)/2$, where the uncertainties
take into account the correlations between $F^{i,n}_+$ and $F^{i,n}_-$.

\begin{table}[h]
\begin{tabular}{cccccccccccccc}
\hline\hline
$|\mathbf{p'}|^2/(2\pi/L)^2$ & \hspace{1ex} & $F_+^\mathtt{C14}$ & \hspace{1ex} & $F_+^\mathtt{C24}$ & \hspace{1ex} &
$F_+^\mathtt{C54}$  & \hspace{1ex} & $F_+^\mathtt{F23}$ & \hspace{1ex} & $F_+^\mathtt{F43}$ & \hspace{1ex} & $F_+^\mathtt{F63}$ \\
\hline
0 && $1.148(53)$ && $1.126(39)$ && $1.119(39)$ && $1.125(74)$ && $1.117(52)$ && $1.069(62)$ \\
1 && $1.030(50)$ && $1.026(37)$ && $1.023(38)$ && $1.037(68)$ && $1.027(48)$ && $0.993(61)$ \\
2 && $0.926(51)$ && $0.923(38)$ && $0.924(38)$ && $0.922(67)$ && $0.921(47)$ && $0.892(64)$ \\
3 && $0.828(53)$ && $0.843(39)$ && $0.842(39)$ && $0.843(69)$ && $0.845(48)$ && $0.814(64)$ \\
4 && $0.776(51)$ && $0.775(38)$ && $0.772(39)$ && $0.795(70)$ && $0.792(49)$ && $0.761(63)$ \\
5 && $0.693(51)$ && $0.719(38)$ && $0.716(39)$ && $0.754(70)$ && $0.747(50)$ && $0.710(63)$ \\
6 && $0.648(52)$ && $0.673(39)$ && $0.664(39)$ && $0.702(72)$ && $0.700(51)$ && $0.673(64)$ \\
8 && $0.578(56)$ && $0.606(41)$ && $0.610(40)$ && $0.632(75)$ && $0.621(55)$ && $0.624(65)$ \\
9 && $0.549(60)$ && $0.568(45)$ && $0.573(42)$ && $0.604(77)$ && $0.590(59)$ && $0.605(67)$ \\
\hline\hline
\end{tabular}
\caption{\label{tab:Fplus} Lattice results for the form factor $F_+$.}
\end{table}

\begin{table}[h]
\begin{tabular}{cccccccccccccc}
\hline\hline
$|\mathbf{p'}|^2/(2\pi/L)^2$ & \hspace{1ex} & $F_-^\mathtt{C14}$ & \hspace{1ex} & $F_-^\mathtt{C24}$ & \hspace{1ex} &
$F_-^\mathtt{C54}$  & \hspace{1ex} & $F_-^\mathtt{F23}$ & \hspace{1ex} & $F_-^\mathtt{F43}$ & \hspace{1ex} & $F_-^\mathtt{F63}$ \\
\hline
1 && $1.80(12)$ && $1.861(98)$ && $1.874(66)$ && $1.70(18)$ && $1.709(93)$   && $1.755(99)$   \\
2 && $1.60(11)$ && $1.615(91)$ && $1.625(63)$ && $1.52(17)$ && $1.540(89)$   && $1.541(98)$   \\
3 && $1.46(12)$ && $1.502(10)$ && $1.513(71)$ && $1.48(18)$ && $1.433(90)$   && $1.415(99)$   \\
4 && $1.17(11)$ && $1.181(85)$ && $1.265(62)$ && $1.29(17)$ && $1.281(89)$   && $1.270(96)$   \\
5 && $1.07(10)$ && $1.110(86)$ && $1.169(63)$ && $1.15(17)$ && $1.177(88)$   && $1.179(95)$   \\
6 && $1.00(11)$ && $1.046(87)$ && $1.101(65)$ && $1.02(17)$ && $1.079(88)$   && $1.115(96)$   \\
8 && $0.82(11)$ && $0.878(88)$ && $0.915(65)$ && $0.89(18)$ && $0.955(90)$   && $0.965(97)$   \\
9 && $0.80(11)$ && $0.814(90)$ && $0.863(69)$ && $0.84(17)$ && $0.93(10)\nb$ && $0.94(10)\nb$ \\
\hline\hline
\end{tabular}
\caption{\label{tab:Fminus} Lattice results for the form factor $F_-$.}
\end{table}

\begin{table}[h]
\begin{tabular}{cccccccccccccc}
\hline\hline
$|\mathbf{p'}|^2/(2\pi/L)^2$ & \hspace{1ex} & $F_1^\mathtt{C14}$ & \hspace{1ex} & $F_1^\mathtt{C24}$ & \hspace{1ex} &
$F_1^\mathtt{C54}$  & \hspace{1ex} & $F_1^\mathtt{F23}$ & \hspace{1ex} & $F_1^\mathtt{F43}$ & \hspace{1ex} & $F_1^\mathtt{F63}$ \\
\hline
1 && $1.417(64)$ && $1.444(56)$ && $1.448(41)$ && $1.368(98)$ && $1.368(47)$ && $1.374(67)$ \\
2 && $1.263(61)$ && $1.269(53)$ && $1.274(41)$ && $1.220(90)$ && $1.231(43)$ && $1.216(69)$ \\
3 && $1.144(62)$ && $1.172(58)$ && $1.177(44)$ && $1.160(92)$ && $1.139(44)$ && $1.115(69)$ \\
4 && $0.975(56)$ && $0.978(50)$ && $1.018(39)$ && $1.041(90)$ && $1.037(43)$ && $1.016(68)$ \\
5 && $0.879(55)$ && $0.914(50)$ && $0.942(40)$ && $0.951(89)$ && $0.962(42)$ && $0.944(67)$ \\
6 && $0.824(56)$ && $0.860(51)$ && $0.883(41)$ && $0.861(91)$ && $0.890(43)$ && $0.894(67)$ \\
8 && $0.699(59)$ && $0.742(51)$ && $0.763(40)$ && $0.760(95)$ && $0.788(46)$ && $0.795(68)$ \\
9 && $0.672(63)$ && $0.691(52)$ && $0.718(41)$ && $0.722(96)$ && $0.760(52)$ && $0.770(71)$ \\
\hline\hline
\end{tabular}
\caption{\label{tab:F1} Lattice results for the form factor $F_1$.}
\end{table}

\begin{table}[h]
\begin{tabular}{cccccccccccccc}
\hline\hline
$|\mathbf{p'}|^2/(2\pi/L)^2$ & \hspace{1ex} & $F_2^\mathtt{C14}$ & \hspace{1ex} & $F_2^\mathtt{C24}$ & \hspace{1ex} &
$F_2^\mathtt{C54}$  & \hspace{1ex} & $F_2^\mathtt{F23}$ & \hspace{1ex} & $F_2^\mathtt{F43}$ & \hspace{1ex} & $F_2^\mathtt{F63}$ \\
\hline
1 && $-0.387(60)$ && $-0.418(43)$ && $-0.425(29)$ && $-0.332(86)$ && $-0.341(47)$ && $-0.381(37)$ \\
2 && $-0.337(56)$ && $-0.346(38)$ && $-0.350(27)$ && $-0.297(82)$ && $-0.309(46)$ && $-0.325(34)$ \\
3 && $-0.316(59)$ && $-0.330(44)$ && $-0.335(31)$ && $-0.317(86)$ && $-0.294(48)$ && $-0.300(35)$ \\
4 && $-0.199(54)$ && $-0.203(36)$ && $-0.247(27)$ && $-0.247(85)$ && $-0.245(48)$ && $-0.254(34)$ \\
5 && $-0.186(54)$ && $-0.196(37)$ && $-0.226(28)$ && $-0.197(82)$ && $-0.215(48)$ && $-0.235(34)$ \\
6 && $-0.176(55)$ && $-0.187(37)$ && $-0.219(29)$ && $-0.159(82)$ && $-0.190(48)$ && $-0.221(35)$ \\
8 && $-0.121(57)$ && $-0.136(41)$ && $-0.153(31)$ && $-0.128(85)$ && $-0.167(50)$ && $-0.171(36)$ \\
9 && $-0.124(60)$ && $-0.123(43)$ && $-0.145(35)$ && $-0.117(82)$ && $-0.170(59)$ && $-0.165(39)$ \\
\hline\hline
\end{tabular}
\caption{\label{tab:F2} Lattice results for the form factor $F_2$.}
\end{table}

\FloatBarrier
\section{\label{sec:chiralcontinuumextrap}Fits of the form factors as functions of $E_N-m_N$, $m_{u,d}$, and $a$}
\FloatBarrier

\begin{figure}
\includegraphics[width=0.465\linewidth]{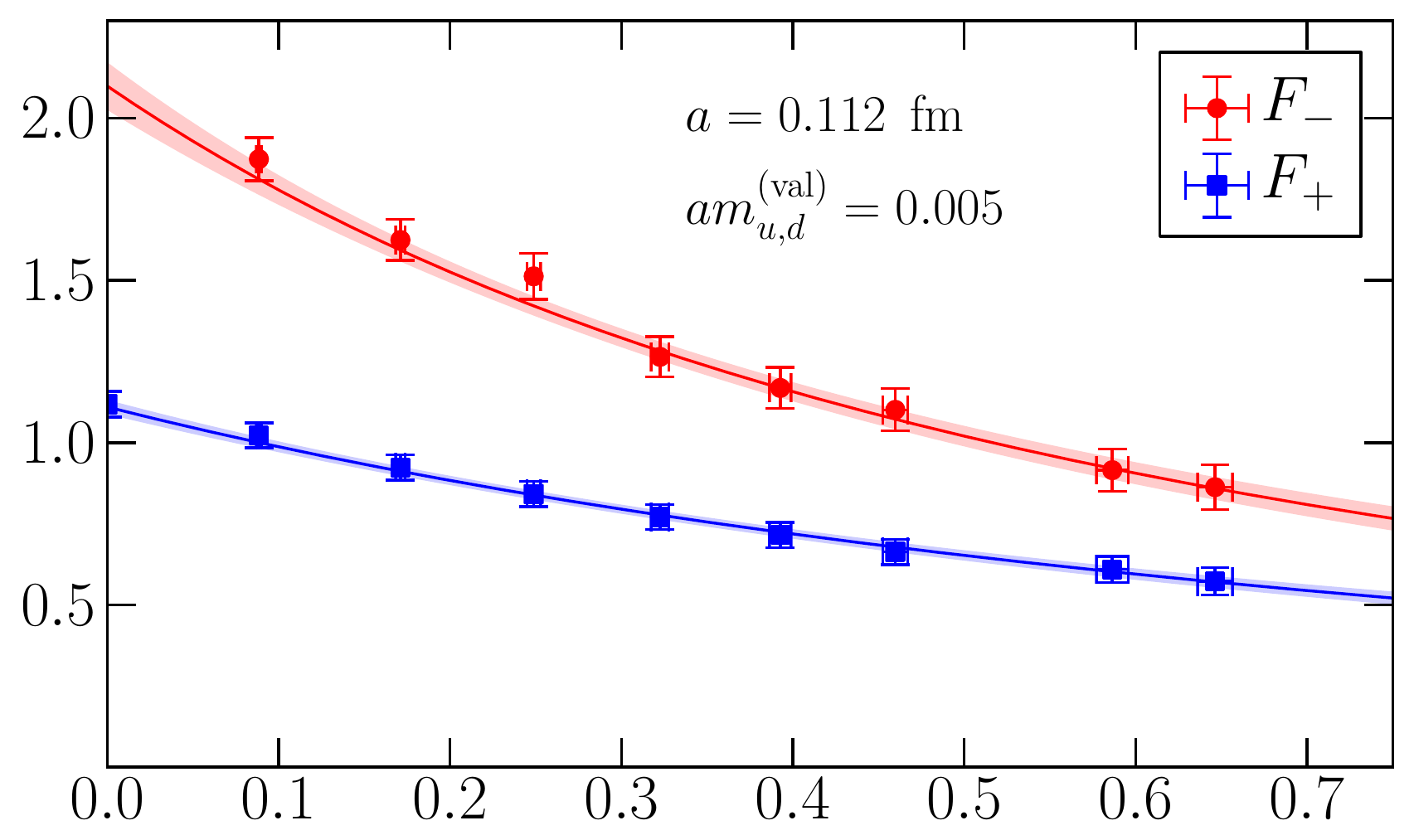}\hfill\includegraphics[width=0.465\linewidth]{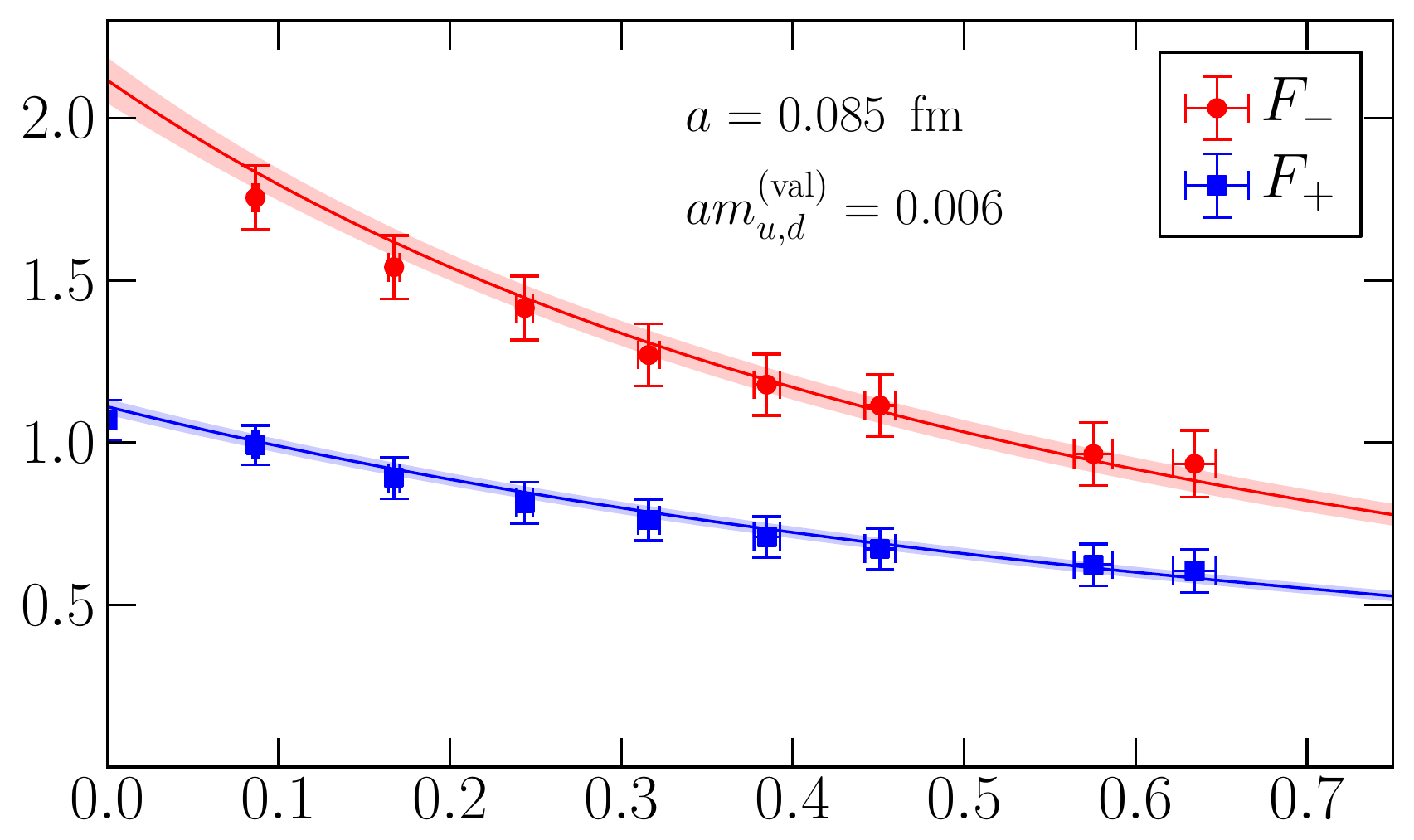} \\
\includegraphics[width=0.465\linewidth]{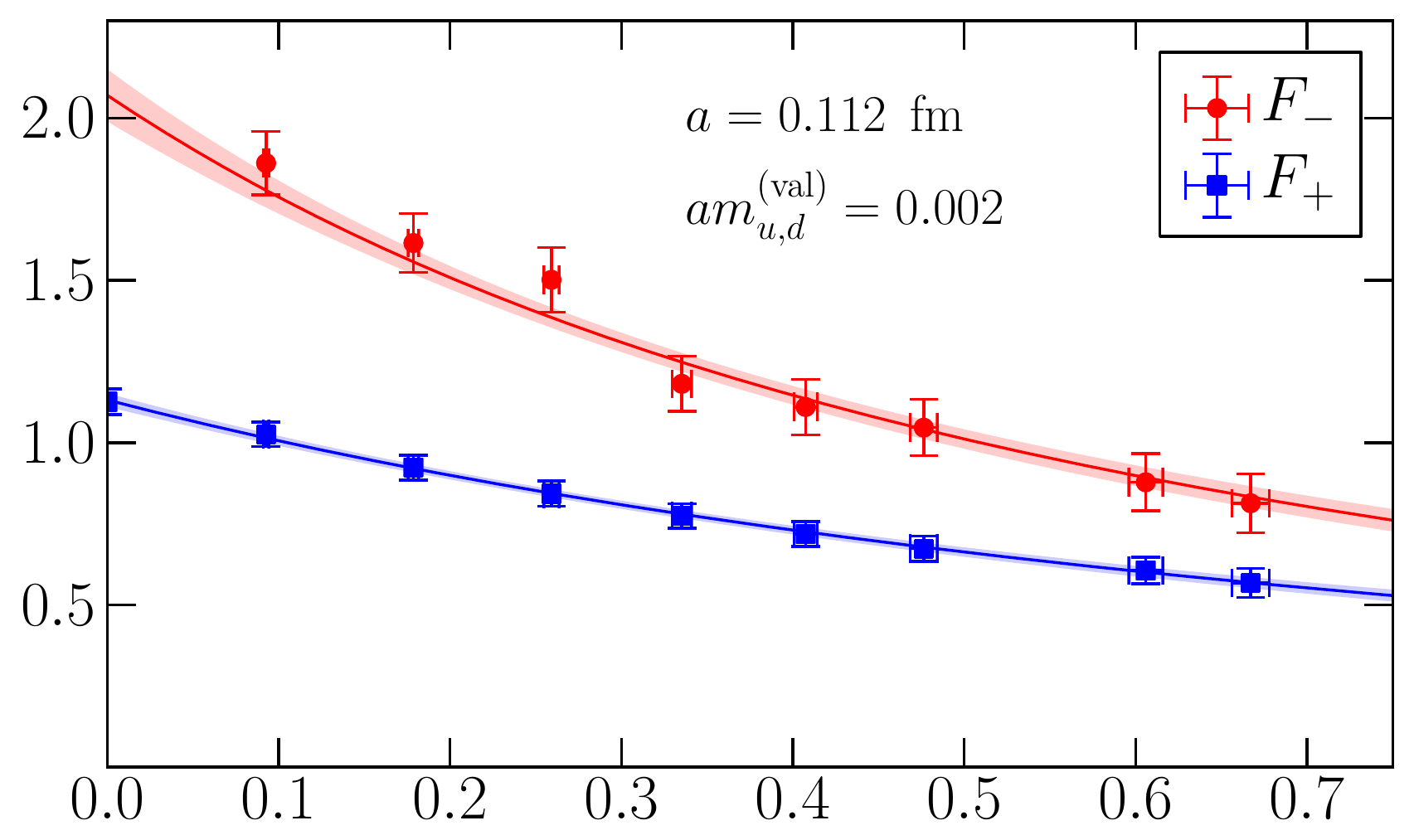}\hfill\includegraphics[width=0.465\linewidth]{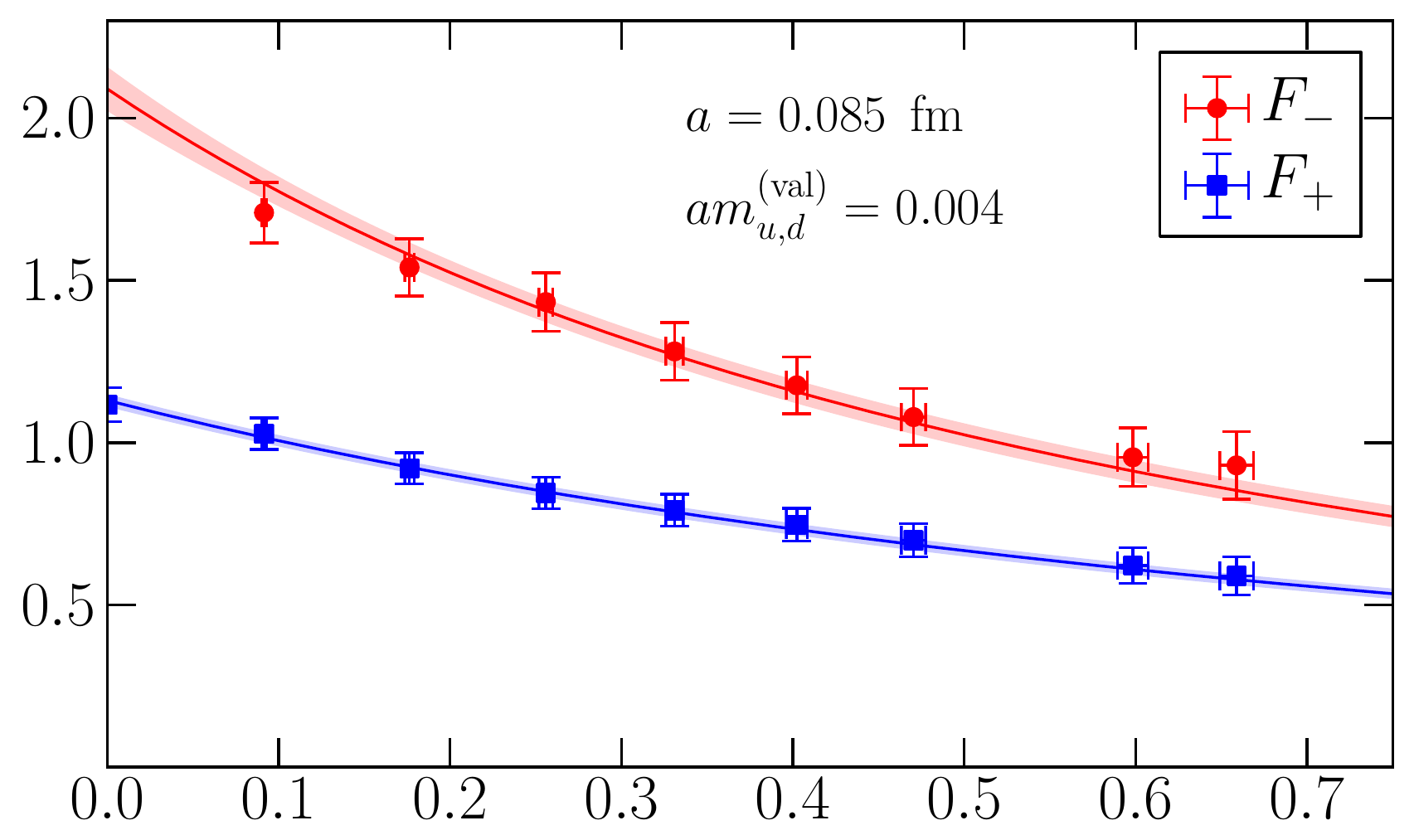} \\
\includegraphics[width=0.465\linewidth]{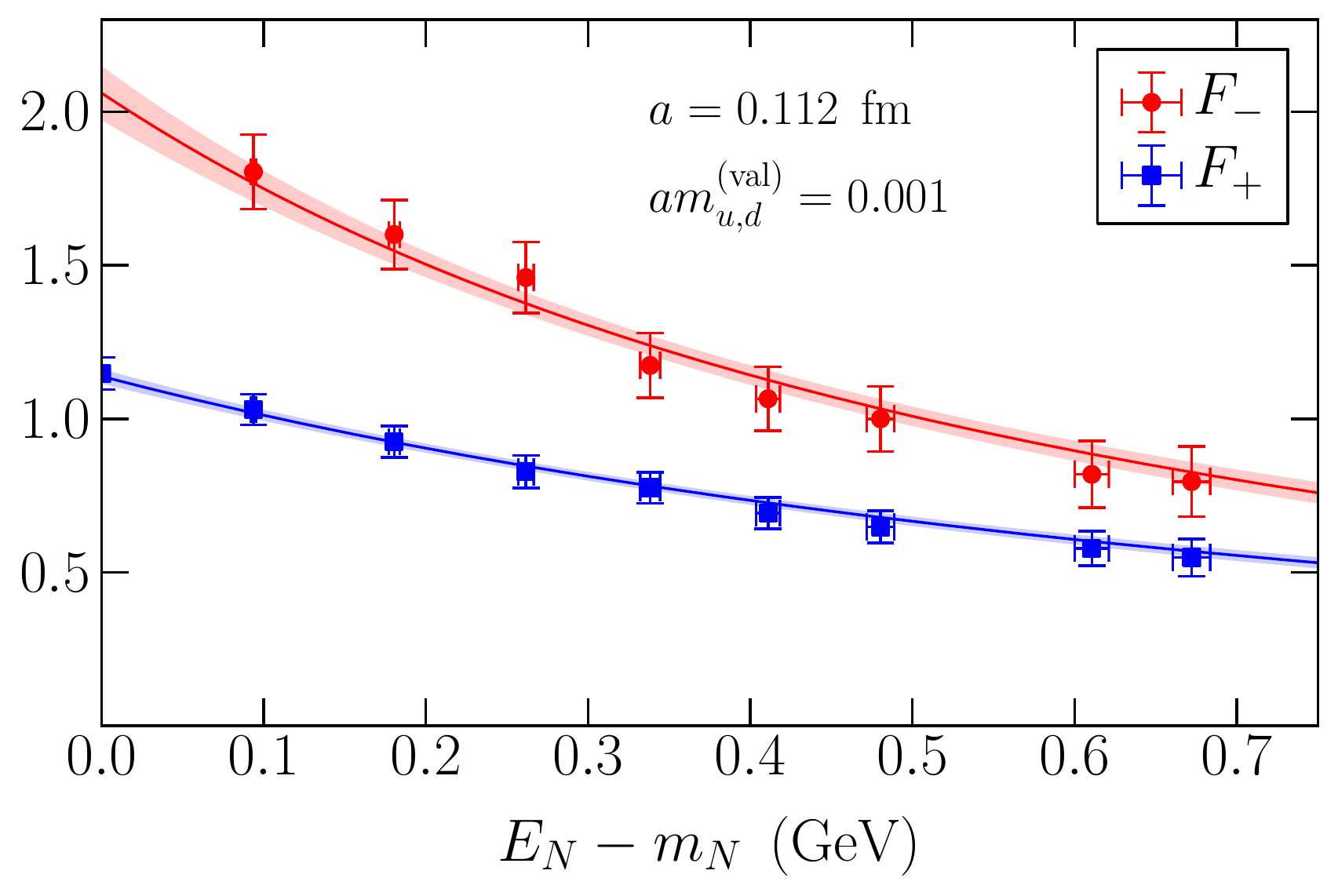}\hfill\includegraphics[width=0.465\linewidth]{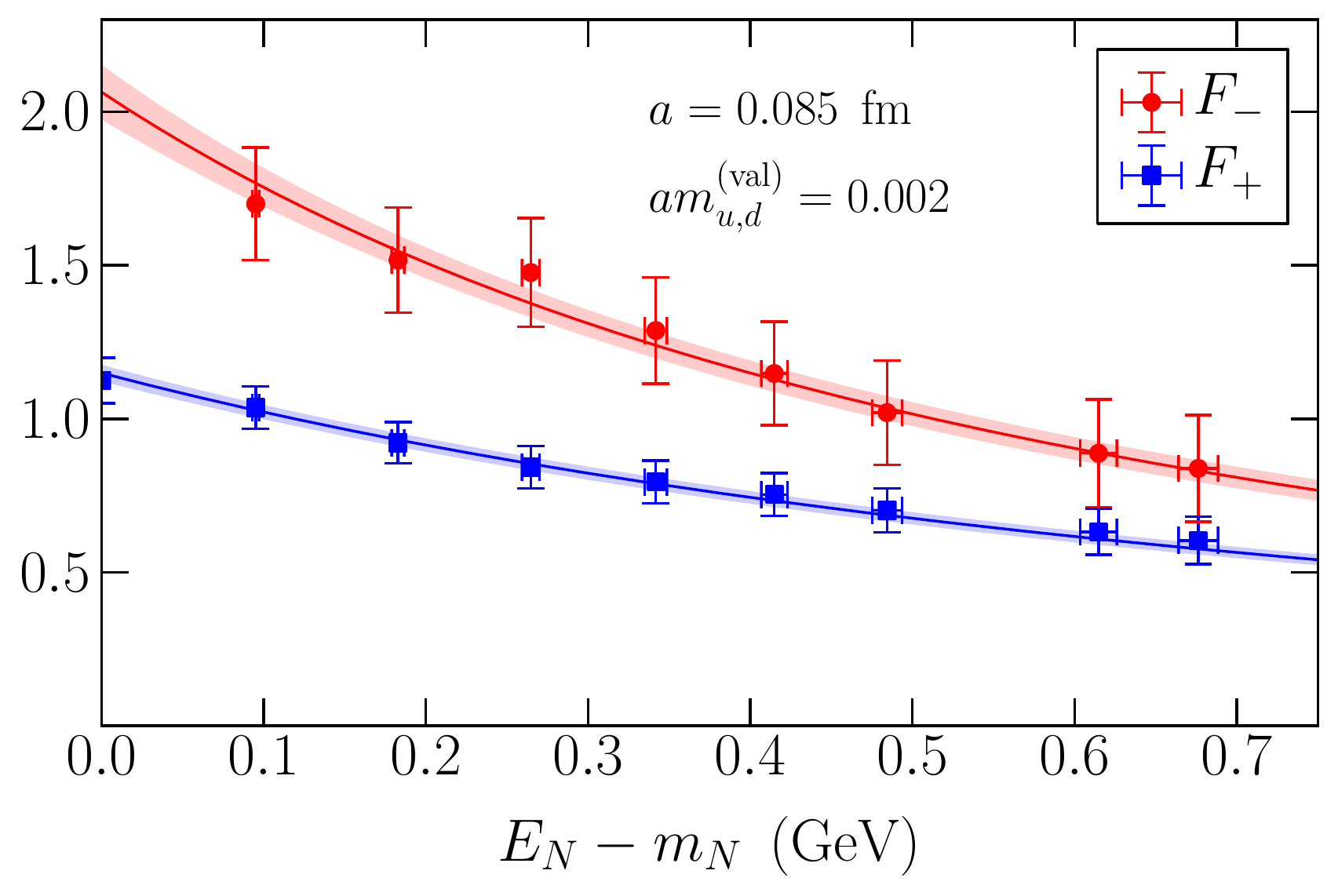} \\
\centerline{\includegraphics[height=5.65cm]{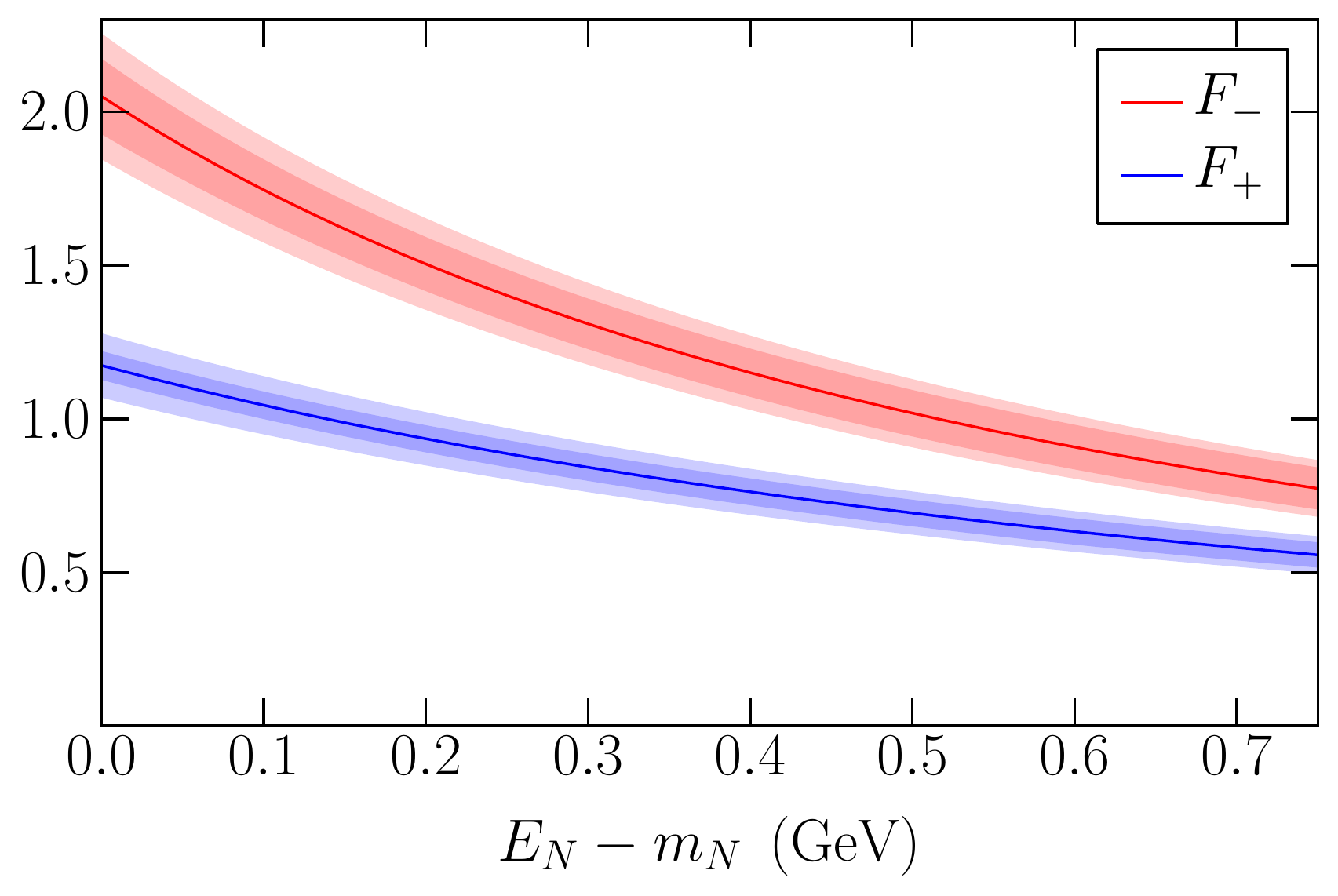}}
\caption{\label{fig:qsqrasqrextrapall}Fits of the form factor data for $F_+$ and $F_-$ using Eq.~(\ref{eq:dipole}).
In the upper six plots, we show the lattice results together with the fitted functions evaluated at the corresponding
values of the pion mass and lattice spacing. In the lower plot, we show the fitted functions evaluated at the
physical pion mass and in the continuum limit. There, the inner shaded bands indicate the statistical/fitting
uncertainty, and the outer shaded bands additionally include the estimates of the systematic uncertainty given
in Eqs.~(\ref{eq:Fplussysterr}) and (\ref{eq:Fminussysterr}).}
\end{figure}

\begin{figure}
\includegraphics[width=0.48\linewidth]{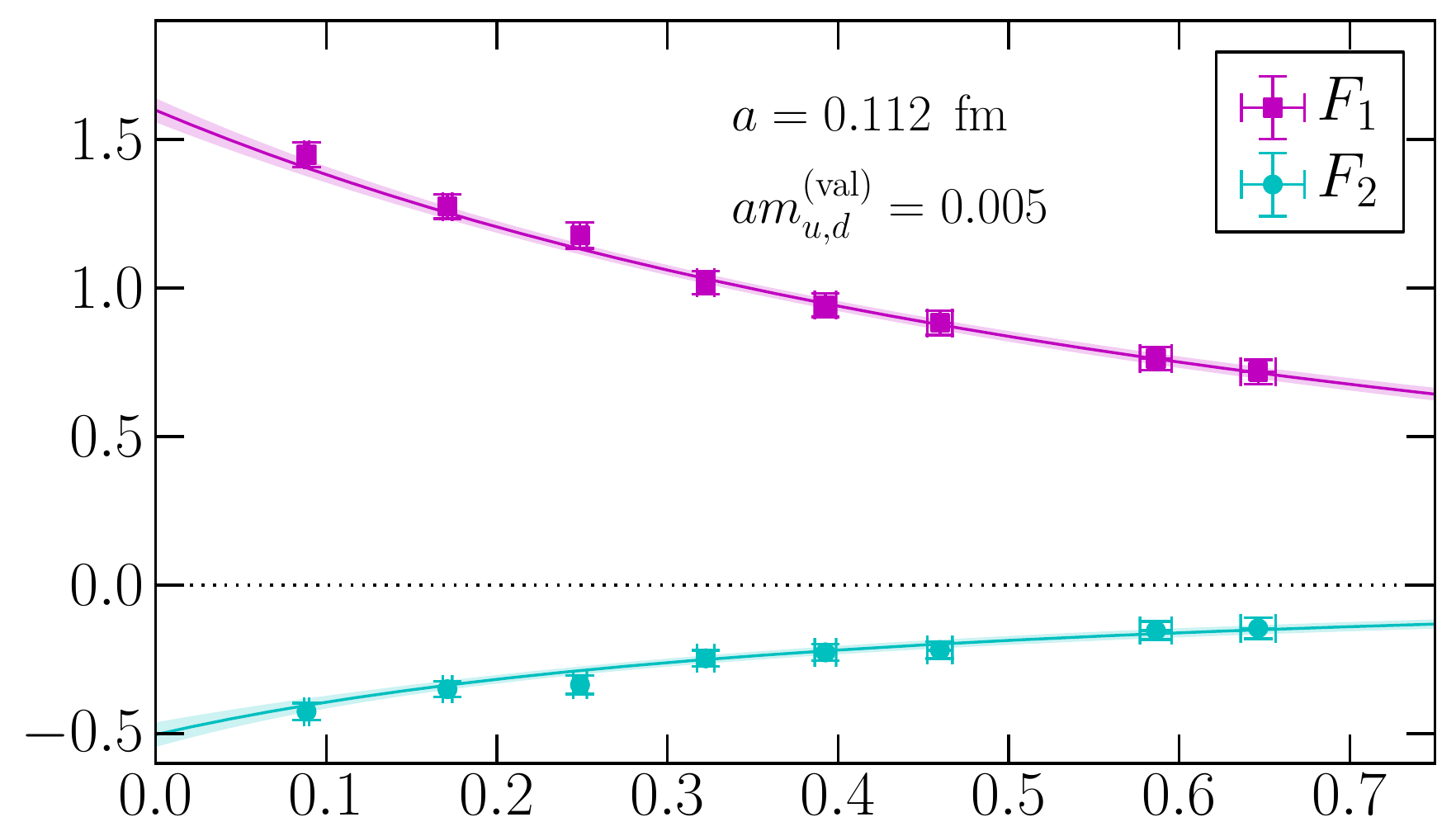}\hfill\includegraphics[width=0.48\linewidth]{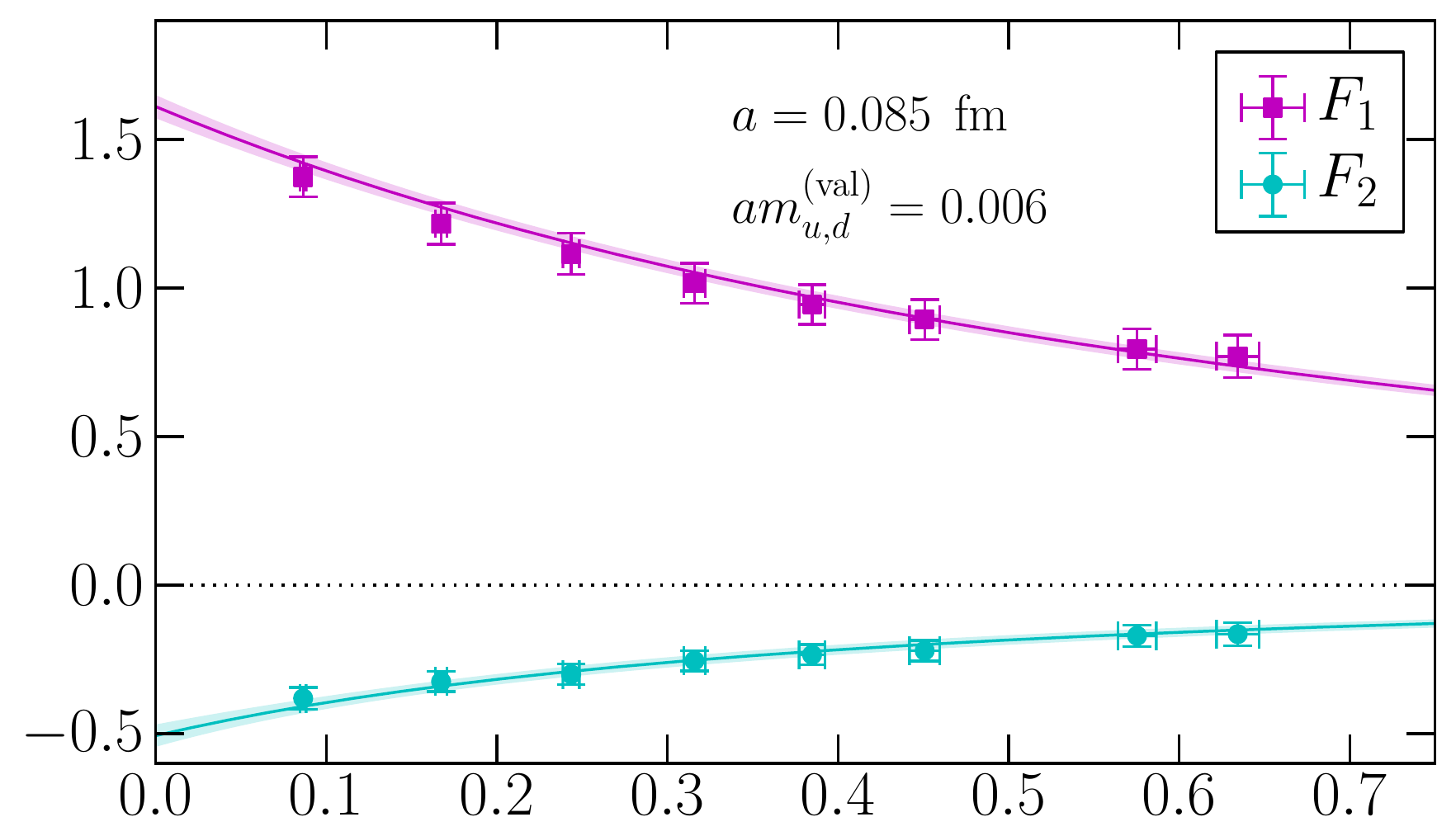} \\
\includegraphics[width=0.48\linewidth]{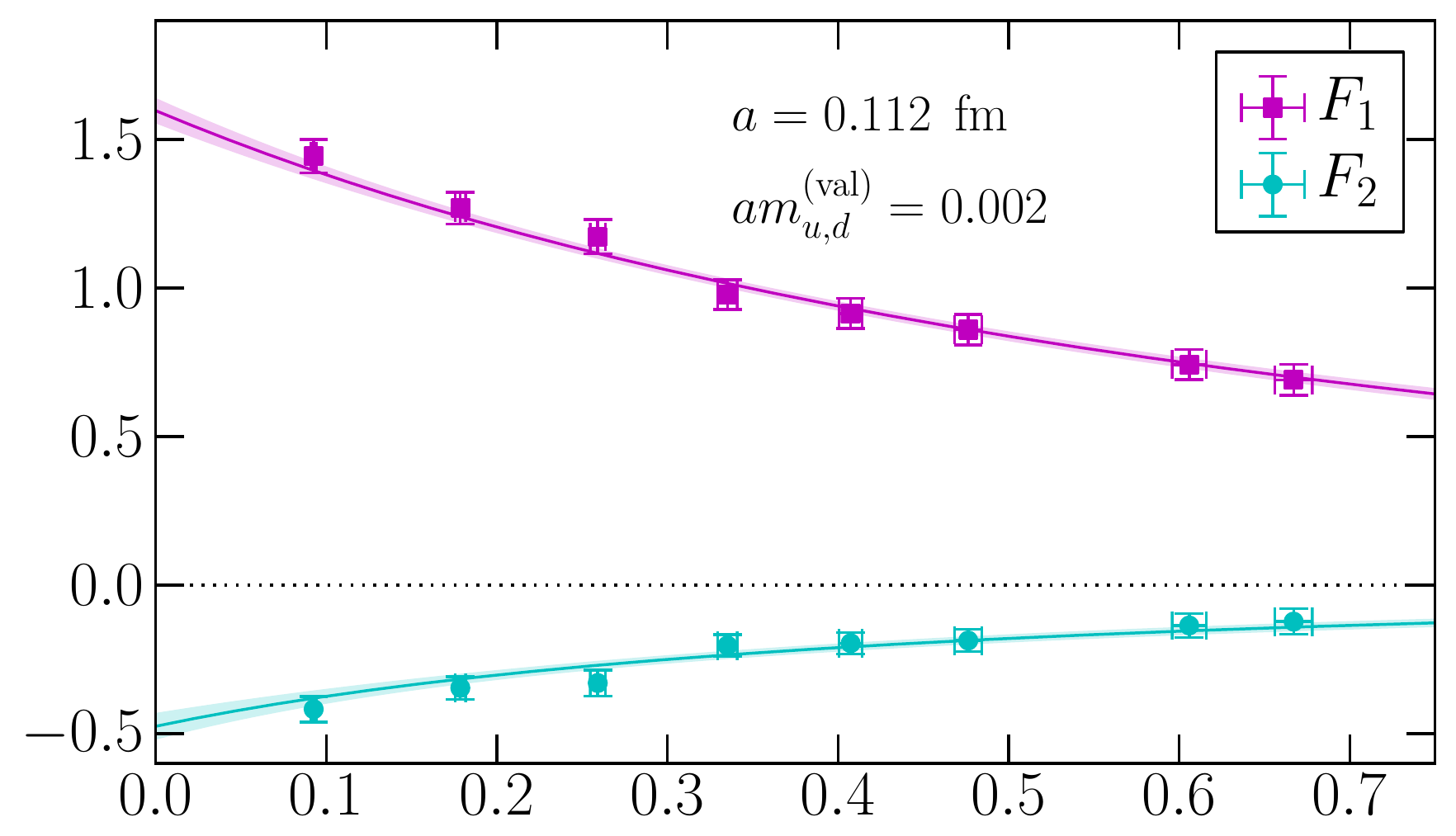}\hfill\includegraphics[width=0.48\linewidth]{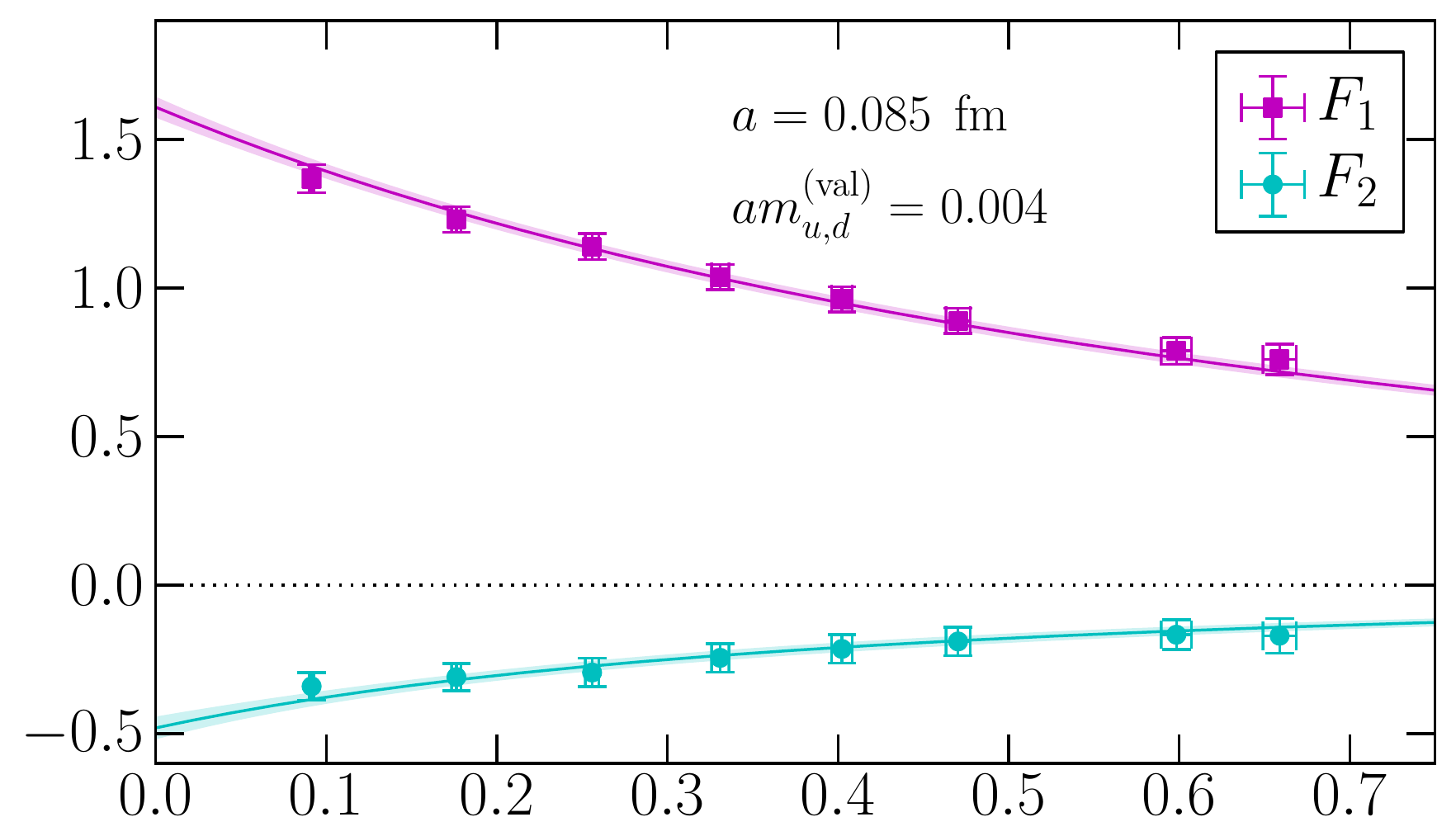} \\
\includegraphics[width=0.48\linewidth]{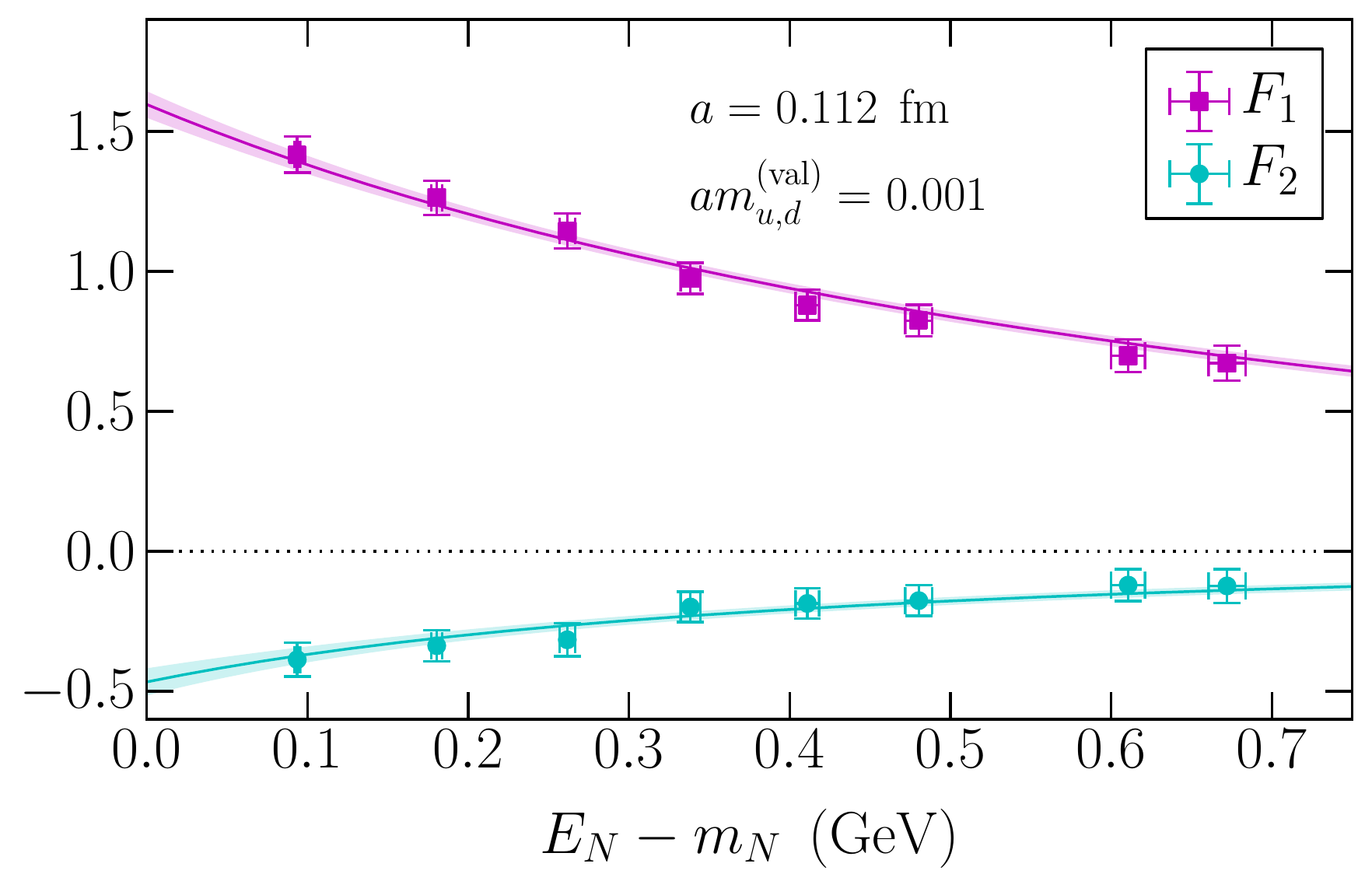}\hfill\includegraphics[width=0.48\linewidth]{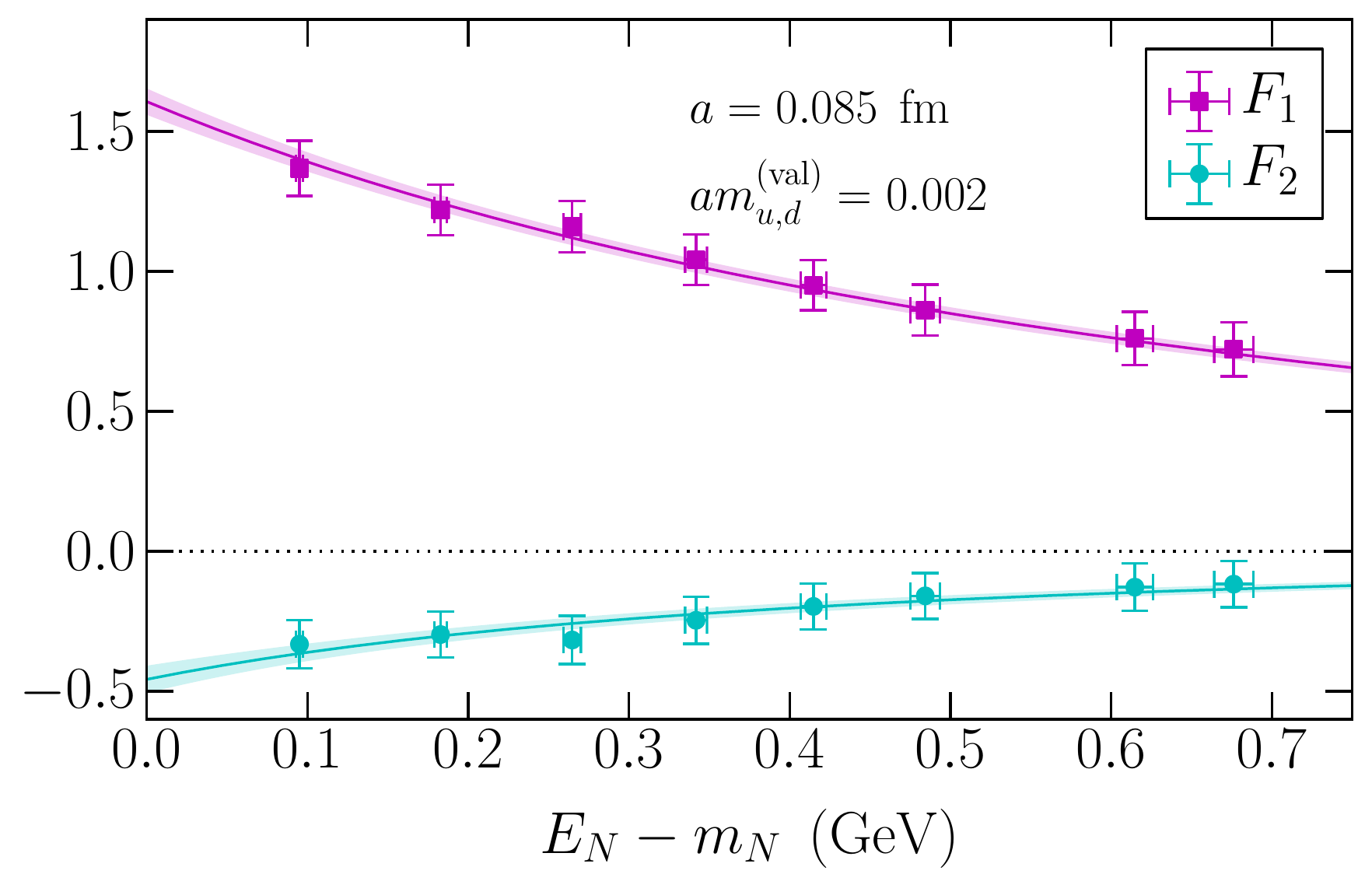} \\
\centerline{\includegraphics[height=5.65cm]{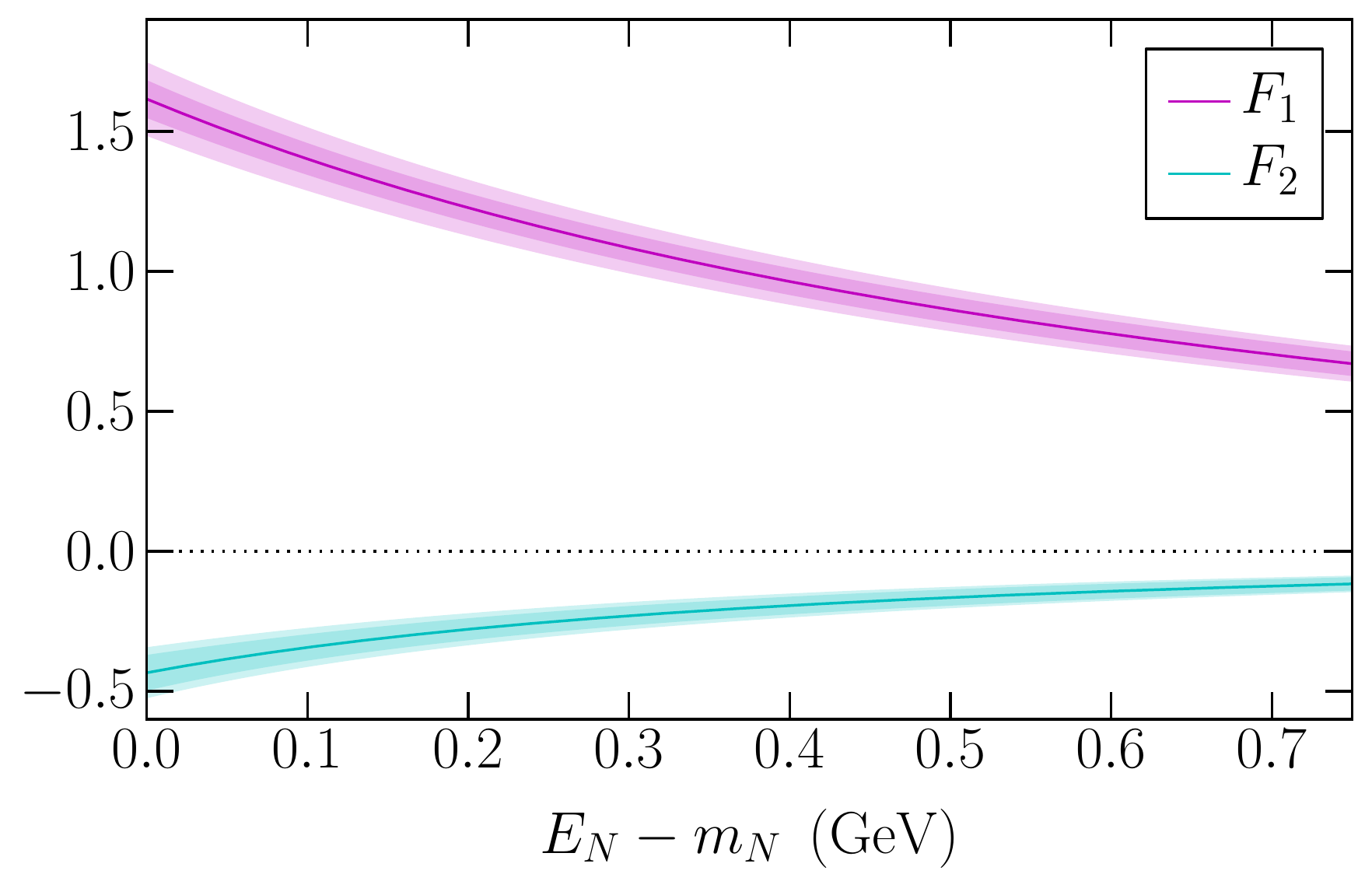}}
\caption{\label{fig:qsqrasqrextrapallF1F2}Fits of the form factor data for $F_1$ and $F_2$ using Eq.~(\ref{eq:dipoleF1F2}).
In the upper six plots, we show the lattice results together with the fitted functions evaluated at the corresponding
values of the pion mass and lattice spacing. In the lower plot, we show the fitted functions evaluated at the physical
pion mass and in the continuum limit. There, the inner shaded bands indicate the statistical/fitting uncertainty,
and the outer shaded bands additionally include the estimates of the systematic uncertainty given in
Eqs.~(\ref{eq:F1systerr}) and (\ref{eq:F2systerr}).}
\end{figure}

In this section we present fits that smoothly interpolate the $E_N$-dependence of our $\Lambda_Q \to p$ form factor
results, including corrections to account for the dependence on the lattice spacing and the light-quark mass. In principle,
the form of this dependence can be predicted in a low-energy effective field theory combining heavy-baryon chiral
perturbation theory for the proton \cite{Jenkins:1991ne, Jenkins:1990jv} with heavy-hadron chiral perturbation theory
\cite{Yan:1992gz, Cho:1992cf} for the $\Lambda_Q$. However, there are a number of issues that limit the usefulness of
this approach for our work. One limitation is that chiral perturbation theory breaks down for momenta $|\mathbf{p'}|$
comparable to or larger than the chiral symmetry breaking scale. Another limitation is that the effective theory also
needs to include the $\Sigma_Q$ and $\Delta$ baryons in the chiral loops, which is expected to lead to additional
unknown low-energy constants associated with the matching of the $Q \to u$ current to the $\Sigma_Q \to p$,
$\Lambda_Q \to \Delta$, and $\Sigma_Q \to \Delta$ currents in the effective theory. Finally, some of the data sets
used here are partially quenched (with valence-quark masses lighter than the sea-quark masses), which further increases
the complexity of the effective theory. As in Ref.~\cite{Detmold:2012vy}, we therefore use a simple model that
successfully describes the dependence of the form factors on $E_N$, $m_{u,d}$, and $a$, at the present level of
uncertainty. It is given by
\begin{eqnarray}
 F_\pm^{i,n} &=& \frac{Y_\pm}{(X_\pm^i+E_N^{i,n}-m_N^i)^2}\cdot [1 + d_\pm (a^i E_N^{i,n})^2], \label{eq:dipole}
\end{eqnarray}
where the position of the pole depends on the pion mass,
\begin{equation}
X_\pm^i = X_\pm + c_\pm \cdot \left[ (m_\pi^i)^2-(m_\pi^{{\rm phys}})^2\right], \label{eq:polemqdep}
\end{equation}
and the term $[1 + d_\pm (a^i E_N^{i,n})^2]$ models the lattice discretization artifacts, which are assumed to
increase with the proton energy. As discussed above, we calculate the proton energies using the relativistic
dispersion relation $E_N^{i,n}=\sqrt{(m_N^i)^2+n\cdot(2\pi/L)^2}$, where $m_N^i$ is the lattice proton mass for
the data set $i$. The free fit parameters in Eq.~(\ref{eq:dipole}) are $Y_{\pm}$, $X_\pm$, $d_\pm$, and $c_\pm$.
Note that here we do not include dependence on the strange-quark mass, because none of the hadrons involved
contain a valence strange quark. The fits of $F_\pm^{i,n}$ using Eq.~(\ref{eq:dipole}) are shown in
Fig.~\ref{fig:qsqrasqrextrapall}, and give the results listed in Table \ref{tab:dipolefitresults}. We have also
performed independent fits of the data for $F^{i,n}_1 = (F^{i,n}_+ + F^{i,n}_-)/2$ and
$F^{i,n}_2=(F^{i,n}_+ - F^{i,n}_-)/2$, using the functions
\begin{eqnarray}
 F_{1,2}^{i,n} &=& \frac{Y_{1,2}}{(X_{1,2}^i+E_N^{i,n}-m_N^i)^2}\cdot [1 + d_{1,2} (a^i E_N^{i,n})^2], \label{eq:dipoleF1F2}
\end{eqnarray}
with $X_{1,2}^i = X_{1,2} + c_{1,2} \cdot \left[ (m_\pi^i)^2-(m_\pi^{{\rm phys}})^2\right]$. These fits
are shown in Fig.~\ref{fig:qsqrasqrextrapallF1F2}, and the resulting values of the parameters are given
in Table \ref{tab:dipolefitresultsF1F2}.

\begin{table}
\begin{tabular}{ccc}
\hline\hline
Parameter & \hspace{1ex} & Result \\
\hline
$Y_+$  && $3.24 \pm 0.62$ ${\rm GeV}^2$  \\
$X_+$  && $1.66 \pm 0.15$ ${\rm GeV}^{\phantom{2}}$  \\
$Y_-$  && $2.92 \pm 0.62$ ${\rm GeV}^2$ \\
$X_-$  && $1.19 \pm 0.13$ ${\rm GeV}^{\phantom{2}}$  \\
\hline\hline
\end{tabular}
\caption{\label{tab:dipolefitresults} Results for the form factor normalization and shape parameters $Y_\pm$ and $X_\pm$
from fits of the lattice QCD results for $F_\pm^{i,n}$, using Eq.~(\protect\ref{eq:dipole}). The covariances of the
parameters needed in Eq.~(\ref{eq:Fplusminusphysical}) are ${\rm Cov}(Y_+,X_+)=0.090\:\:{\rm GeV}^3$ and
${\rm Cov}(Y_-,X_-)=0.080\:\:{\rm GeV}^3$. The results for the parameters describing the quark mass and lattice
spacing dependence are $c_+=0.38(35)\:\:{\rm GeV}^{-1}$, $d_+=-0.031(81)$, $c_-=-0.22(35)\:\:{\rm GeV}^{-1}$,
and $d_-=-0.025(94)$.}
\end{table}

\begin{table}
\begin{tabular}{ccc}
\hline\hline
Parameter & \hspace{1ex} & Result \\
\hline
$Y_1$  && $\wm2.97 \pm 0.50$ ${\rm GeV}^2$  \\
$X_1$  && $\wm1.36 \pm 0.12$ ${\rm GeV}^{\phantom{2}}$  \\
$Y_2$  && $  -0.28 \pm 0.11$   ${\rm GeV}^2$ \\
$X_2$  && $\wm0.81 \pm 0.17$ ${\rm GeV}^{\phantom{2}}$  \\
\hline\hline
\end{tabular}
\caption{\label{tab:dipolefitresultsF1F2} Results for the form factor normalization and shape parameters $Y_{1,2}$ and
$X_{1,2}$ from fits of the lattice QCD results for $F_{1,2}^{i,n}$, using Eq.~(\protect\ref{eq:dipoleF1F2}).
The covariances of the parameters needed in Eq.~(\ref{eq:F12physical}) are
${\rm Cov}(Y_1,X_1)=0.057\:\:{\rm GeV}^3$ and ${\rm Cov}(Y_2,X_2)=-0.018\:\:{\rm GeV}^3$.
The results for the parameters describing the quark mass and lattice spacing dependence are $d_1=-0.038(70)$,
$c_1=-0.04(30)\:\:{\rm GeV}^{-1}$, $d_2=0.05(22)$, and $c_2=-0.53(54)\:\:{\rm GeV}^{-1}$.}
\end{table}

By construction, Eqs.~(\ref{eq:dipole}) and (\ref{eq:dipoleF1F2}) reduce to
\begin{eqnarray}
 F_\pm &=& \frac{Y_\pm}{(X_\pm+E_N-m_N)^2} \label{eq:Fplusminusphysical}, \\
 F_{1,2} &=& \frac{Y_{1,2}}{(X_{1,2}+E_N-m_N)^2} \label{eq:F12physical}
\end{eqnarray}
in the continuum limit and at the physical pion mass. These functions are shown at the bottom of Figs.~\ref{fig:qsqrasqrextrapall}
and \ref{fig:qsqrasqrextrapallF1F2}. In the range of $E_N-m_N$ considered here, the numerical results for
Eqs.~(\ref{eq:Fplusminusphysical}) and (\ref{eq:F12physical}) are consistent with the relations $F_+=F_1+F_2$ and
$F_-=F_1-F_2$ within the statistical uncertainties, as expected. In the plots at the bottom of Figs.~\ref{fig:qsqrasqrextrapall}
and \ref{fig:qsqrasqrextrapallF1F2}, the statistical/fitting uncertainty is indicated using the inner error bands. The
outer error bands additionally include estimates of the total systematic uncertainty, arising from the following sources:
the matching of the lattice HQET to continuum HQET current, the finite lattice volume, the unphysical light-quark
masses, and the non-zero lattice spacing. We discuss these uncertainties below.

As explained in Sec.~\ref{sec:latticecalc}, the lattice HQET to continuum HQET matching is performed using one-loop
perturbation theory at the scale $\mu=a^{-1}$, followed by a two-loop renormalization-group evolution from $\mu=a^{-1}$
to the scale of the $b$-quark mass. To estimate the uncertainty resulting from this use of perturbation theory, we vary
the scale from $\mu=a^{-1}$ to $\mu=2a^{-1}$. For the $\Lambda_Q \to p$ form factors, this results in a change by 7\%
at the coarse lattice spacing, and 6\% at the fine lattice spacing (these relative changes are approximately the same
as for $\Lambda_Q \to \Lambda$ \cite{Detmold:2012vy}; any difference in the size of the effect has to come from the
$\mathcal{O}(a)$-improvement terms, but their contribution is small). Thus, we take the matching uncertainty for the
continuum-extrapolated form factors to be 6\%. Finite-volume effects are also estimated in the same way as in
Ref.~\cite{Detmold:2012vy}; based on the values of $\exp(-m_\pi L)$ for each data set we estimate the finite-volume
effects in the extrapolated form factors to be of order 3\%. The extrapolations to the physical pion mass and the
continuum limit using our simple fit models (\ref{eq:dipole}) and (\ref{eq:dipoleF1F2}) with a small number of
parameters cannot be expected to completely remove the errors associated with the unphysical light-quark masses
and nonzero lattice spacing. As discussed above, we did not use chiral perturbation theory, and we ignored the fact
that some of our lattice results are partially quenched. Similarly, our fit models assume a particular $E_N$-dependence
of the lattice-spacing errors, which was not derived from effective field theory. Following Ref.~\cite{Detmold:2012vy},
we estimate the resulting systematic uncertainties by comparing the form factor results from our standard fits to those
from fits with the parameters $c_\pm$, $c_{1,2}$ or $d_\pm$, $d_{1,2}$ set to zero. In the energy range
$0\leq E_N-m_N\leq 0.7$ GeV, the maximum changes when setting $c_\pm=0$, $c_{1,2}=0$ are 3\% for $F_+$, 3\% for $F_-$,
1\% for $F_1$, and 13\% for $F_2$. In the same range, the maximum changes when setting $d_\pm=0$, $d_{1,2}=0$ are 2\% for
$F_+$, 2\% for $F_-$, 3\% for $F_1$, and 4\% for $F_2$. None of these changes are statistically significant; nevertheless
we add these percentages in quadrature to the uncertainties from the current matching and from the finite-volume effects.

In summary, we obtain the following estimates of the total systematic uncertainties (valid for $0\leq E_N-m_N\leq 0.7$ GeV):
\begin{eqnarray}
 F_+: && \hspace{4ex} \sqrt{(6\%)^2 + (3\%)^2 + (3\%)^2 + (2\%)^2 } \approx 8\%, \label{eq:Fplussysterr} \\
 F_-: && \hspace{4ex} \sqrt{(6\%)^2 + (3\%)^2 + (3\%)^2 + (2\%)^2 } \approx 8\%, \label{eq:Fminussysterr} \\
 F_1: && \hspace{4ex} \sqrt{(6\%)^2 + (3\%)^2 + (1\%)^2 + (3\%)^2 } \approx 7\%, \label{eq:F1systerr} \\
 F_2: && \hspace{4ex} \sqrt{(6\%)^2 + (3\%)^2 + (13\%)^2 + (4\%)^2 } \approx 15\% \label{eq:F2systerr}.
\end{eqnarray}
Note that in contrast to Ref.~\cite{Detmold:2012vy}, here we choose to only evaluate the form factors in the
energy region where we have lattice data, and Eqs.~(\ref{eq:dipole}) and (\ref{eq:dipoleF1F2}) interpolate this data
in $E_N-m_N$. While we have investigated extrapolations into the large-energy region, such extrapolations necessarily
introduce model dependence (similar to that seen in Ref.~\cite{Detmold:2012vy}) and will not aid in a precision
extraction of $|V_{ub}|$ from experiment.

\FloatBarrier
\section{\label{sec:FFcomparison}Comparison with other form factor results}
\FloatBarrier

\begin{figure}[!t]
\includegraphics[height=7cm]{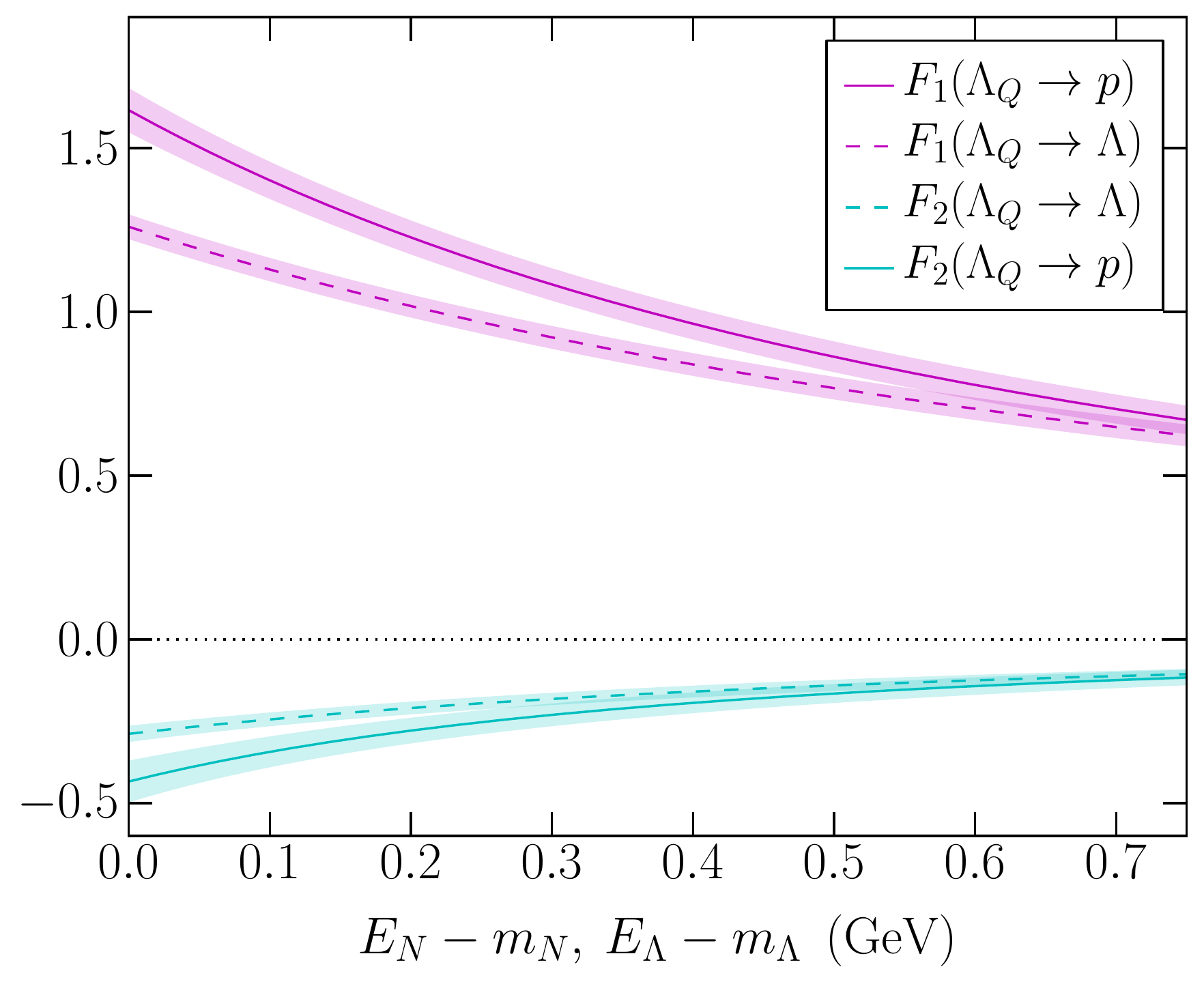} \hfill \includegraphics[height=7cm]{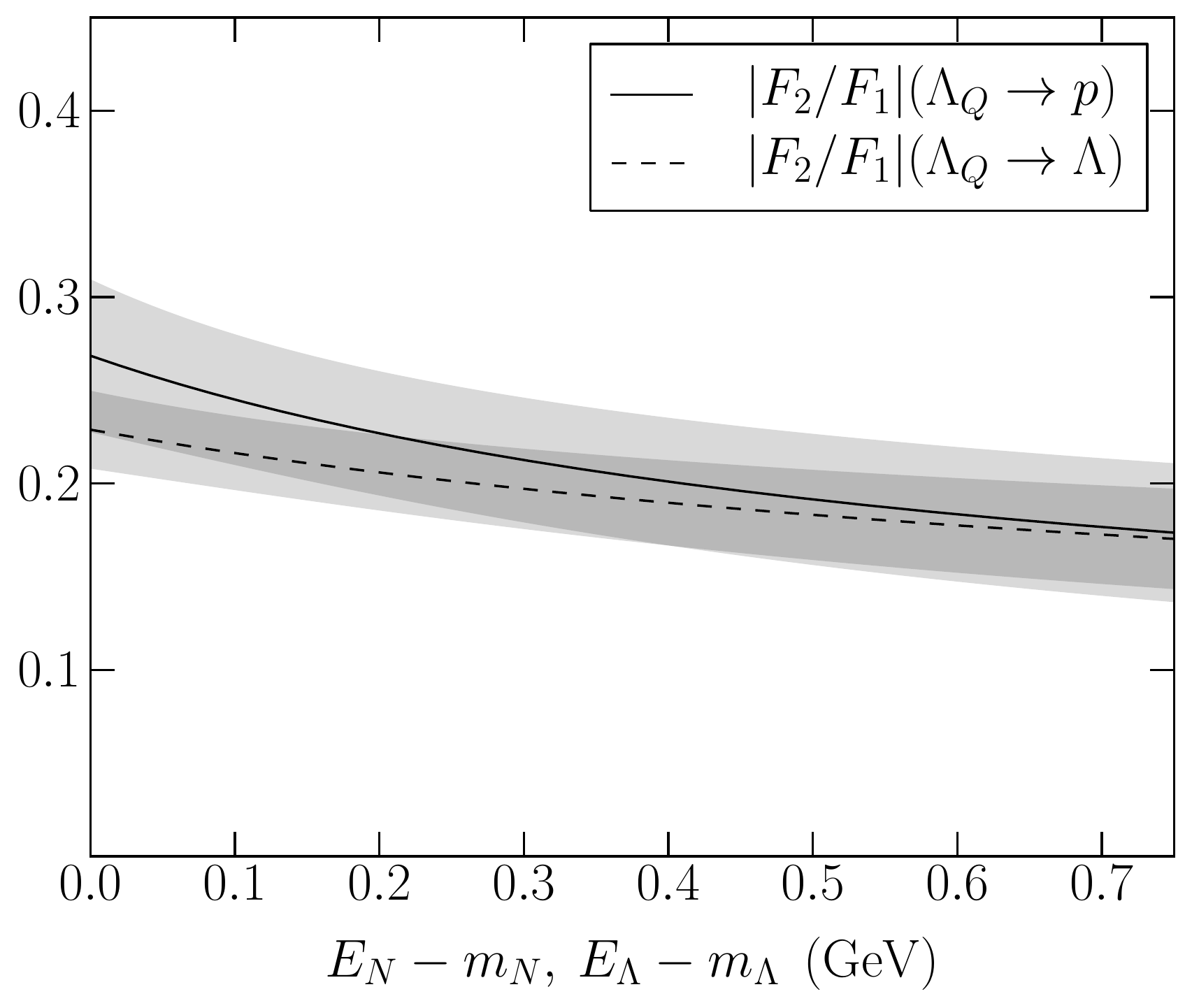}
\caption{\label{fig:protonvsLambda}Left: comparison of the form factors $F_1$ and $F_2$ for the $\Lambda_Q \to p$
transition to the analogous form factors for the $\Lambda_Q \to \Lambda$ transition \cite{Detmold:2012vy}, all
calculated using the same actions and parameters in lattice QCD. Right: comparison of the ratio $|F_2/F_1|$.
Only the statistical error bands are shown here for clarity.}
\end{figure}

It is interesting to compare our results for the $\Lambda_Q \to p$ form form factors to the corresponding results
for the $\Lambda_Q \to \Lambda$ transition obtained in Ref.~\cite{Detmold:2012vy}. This comparison is shown for
$F_1$ and $F_2$ in Fig.~\ref{fig:protonvsLambda}, where we plot the form factors vs.~$E_N - m_N$ and
$E_\Lambda - m_\Lambda$ as before. This choice of variables on the horizontal axis ensures that the points of
zero spatial momentum of the final-state hadron (in the $\Lambda_Q$ rest frame) coincide. As can be seen in the
figure, when compared in this way, the $\Lambda_Q \to p$ form factors have a larger magnitude than the
$\Lambda_Q \to \Lambda$ form factors. This difference is statistically most significant at zero recoil, and becomes
less well resolved at higher energy, where our relative uncertainties grow. For the ratio $F_2/F_1$, we are unable
to resolve any difference between $\Lambda_Q \to p$ and $\Lambda_Q \to \Lambda$, as shown on the right-hand side
of Fig.~\ref{fig:protonvsLambda}.

It is also interesting to compare our QCD calculation of the $\Lambda_Q \to p$ form form factors with calculations
using sum rules \cite{Huang:1998rq, Carvalho:1999ia} or light-cone sum rules
\cite{Huang:2004vf, Wang:2009hra, Azizi:2009wn, Khodjamirian:2011jp}. However, most of these studies worked with the
relativistic form factors, and focused on the region of high proton momentum (low $q^2$), where our results would involve
extrapolation and hence model dependence. Only Ref.~\cite{Huang:1998rq} explicitly includes results for the HQET form
factors $F_1$ and $F_2$ in an energy region that overlaps with the region where we have lattice data. For example,
at $E_N-m_N=0.7$ GeV, the results obtained in Ref.~\cite{Huang:1998rq} for three different values of the Borel
parameter used in that work are $F_1\approx (0.46, 0.47, 0.50)$ and $F_2\approx (-0.13, -0.18, -0.27)$, while our
lattice QCD calculation gives
\begin{eqnarray}
 F_1(E_N-m_N=0.7\:{\rm GeV}) &=& 0.703 \pm 0.045 \pm 0.049,  \\
 F_2(E_N-m_N=0.7\:{\rm GeV}) &=& -0.124 \pm 0.025 \pm 0.019,
\end{eqnarray}
where the first uncertainty is statistical and the second uncertainty is systematic.

\section{\label{sec:Lambdabdecay}The decay $\Lambda_b \to p\: \ell^-\, \bar{\nu}_\ell$}

In this section, we use the form factors determined above to calculate the differential decay rates of
$\Lambda_b \to p\: \ell^-\, \bar{\nu_\ell}$ with $\ell=e,\mu,\tau$ in the Standard Model. The effective weak Hamiltonian
for $b \to u \:\ell^-\, \bar{\nu}_\ell$ transitions is
\begin{equation}
\mathcal{H}_{\rm eff} = \frac{G_F}{\sqrt{2}}V_{ub}\: \bar{u} \gamma_\mu(1-\gamma_5)b\: \bar{l} \gamma^\mu (1-\gamma_5) \nu,
\end{equation}
with the Fermi constant $G_F$ and the CKM matrix element $V_{ub}$ \cite{Cabibbo:1963yz, Kobayashi:1973fv, Wilson:1969zs}.
Higher-order electroweak corrections are neglected. The resulting amplitude for the decay
$\Lambda_b \to p\: \ell^-\, \bar{\nu}_\ell$ can be written as
\begin{eqnarray}
\nonumber \mathcal{M}&=&-i\,\langle N^+(p',s')\:\ell^-(p_-,s_-)\:\bar{\nu}(p_+,s_+) | \mathcal{H}_{\rm eff} | \Lambda_b(p,s) \rangle \\
 &=& -i \frac{G_F}{\sqrt{2}}V_{ub}\: A_\mu \:\bar{u}_{\ell}(p_-,s_-)\gamma^\mu(1-\gamma_5) v_{\bar{\nu}}(p_+,s_+),
\end{eqnarray}
where $A_\mu$ is the hadronic matrix element
\begin{equation}
A_\mu = \langle N^+(p',s')| \: \bar{u} \gamma_\mu(1-\gamma_5)b\: | \Lambda_b(p,s) \rangle.  \label{eq:hadMEqcd}
\end{equation}
Because we have computed the form factors in HQET, we need to match the QCD current $\bar{u} \gamma_\mu(1-\gamma_5)b$ in
Eq.~(\ref{eq:hadMEqcd}) to the effective theory. This gives (at leading order in $1/m_b$)
\begin{equation}
A_\mu = \sqrt{m_{\Lambda_b}} \langle N^+(p',s')| \big( c_\gamma\: \bar{u} \gamma_\mu Q + c_v\: \bar{u} v_\mu Q
- c_\gamma\: \bar{u} \gamma_\mu\gamma_5 Q + c_v\: \bar{u} v_\mu \gamma_5 Q  \big) | \Lambda_Q(v,s) \rangle,  \label{eq:hadMEHQET1}
\end{equation}
where $Q$ is the static heavy-quark field, and to one loop, the matching coefficients are given by \cite{Eichten:1989zv}
\begin{eqnarray}
c_\gamma &=& 1-\frac{\alpha_s(\mu)}{\pi}\left[ \frac{4}{3} + \ln\left( \frac{\mu}{m_b} \right) \right], \\
c_v      &=& \frac{2}{3} \frac{\alpha_s(\mu)}{\pi}.
\end{eqnarray}
Here we set $\mu=m_b$. We can now use Eq.~(\ref{eq:FFdef}) to express the matrix element $A_\mu$ in terms of the form
factors $F_1$ and $F_2$:
\begin{equation}
A_\mu = \bar{u}_N(p',s')\Big(F_1+\slashed{v} F_2\Big)\Big( c_\gamma \gamma_\mu + c_v \: v_\mu 
- c_\gamma \: \gamma_\mu\gamma_5 +  c_v\: v_\mu\gamma_5  \Big) \sqrt{m_{\Lambda_b}} u_{\Lambda_Q}(v,s). \label{eq:hadMEHQET2}
\end{equation}
The factor of $\sqrt{m_{\Lambda_b}}$ in Eqs.~(\ref{eq:hadMEHQET1}) and (\ref{eq:hadMEHQET2}) results from the HQET
convention for the normalization of the state $| \Lambda_Q(v,s) \rangle$ and the spinor $u_{\Lambda_Q}(v,s)$. We can make
the replacement $\sqrt{m_{\Lambda_b}}\:u_{\Lambda_Q}(v, s) = u_{\Lambda_b}(p,s)$, where $p=m_{\Lambda_b} v$, and the
spinor $u_{\Lambda_b}(p,s)$ has the standard relativistic normalization. A straightforward calculation then gives the
following differential decay rate,
\begin{eqnarray}
\nonumber \frac{d\Gamma}{d q^2} &=& \frac{|V_{ub}|^2 G_F^2}{768 \pi^3 q^6 m_{\Lambda_b}^5} (q^2 - m_\ell^2)^2
\sqrt{((m_{\Lambda_b}+m_N)^2-q^2)((m_{\Lambda_b}-m_N)^2-q^2)}\\
&& \times \Bigg[  (4 c_\gamma^2+4 c_\gamma c_v+2c_v^2) m_\ell^2 \mathcal{F}\, \mathcal{I}+
\Big( 2 c_\gamma^2 (\mathcal{I}+3 q^2 m_{\Lambda_b}^2) + c_v (2 c_\gamma+c_v)(\mathcal{I}- 3 q^2 m_{\Lambda_b}^2) \Big)
q^2\mathcal{F} +4 c_\gamma (c_{\gamma}+c_v) \mathcal{K} \Bigg], \hspace{3ex}
\end{eqnarray}
where we have defined the combinations
\begin{eqnarray}
 \mathcal{F}&=&((m_{\Lambda_b}+m_N)^2-q^2)F_+^2+((m_{\Lambda_b}-m_N)^2-q^2) F_-^2, \\
 \mathcal{I} &=& m_{\Lambda_b}^4-2 m_N^2 (m_{\Lambda_b}^2+q^2)+q^2 m_{\Lambda_b}^2+m_N^4+q^4, \\
 \mathcal{K} &=& (2 m_\ell^2+q^2) ((m_{\Lambda_b}+m_N)^2-q^2)((m_{\Lambda_b}-m_N)^2-q^2)
   (m_{\Lambda_b}^2-m_N^2+q^2)  F_+ F_-,
\end{eqnarray}
and, as before, $F_\pm=F_1\pm F_2$. To evaluate this, we use $F_+$ and $F_-$ from the fits to our lattice QCD results,
which are parametrized by Eq.~(\ref{eq:dipole}) with the parameters in Table \ref{tab:dipolefitresults}. At a given
value of $q^2$, we evaluate the form factors at
\begin{equation}
 E_{N}-m_N = p'\cdot v-m_N = \frac{m_{\Lambda_b}^2+m_N^2-q^2}{2m_{\Lambda_b}}-m_N, \label{eq:E_N}
\end{equation}
with the physical values of the baryon masses (which we take from Ref.~\cite{PDG2012}).

\begin{figure}
\includegraphics[height=6.5cm]{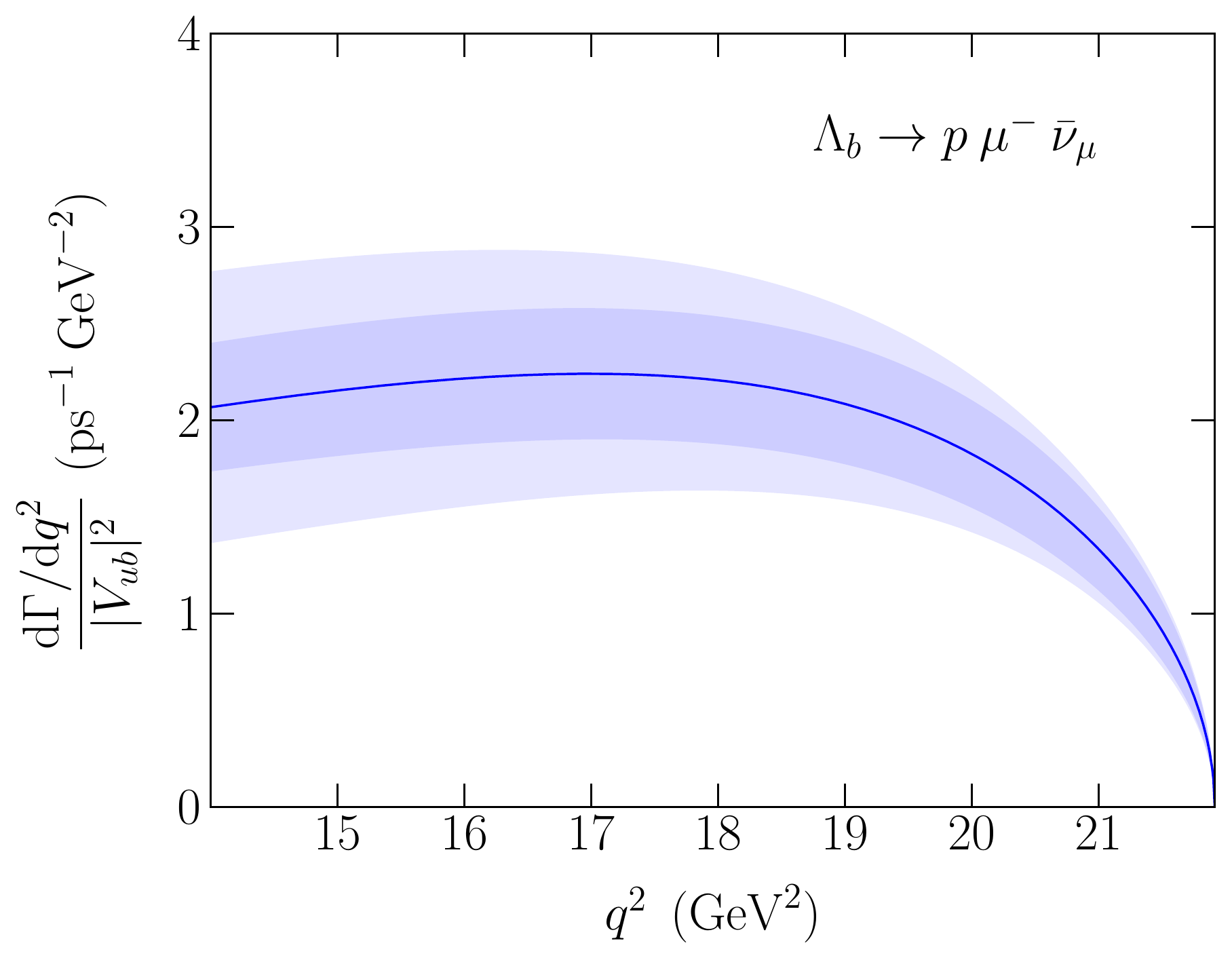} \hfill \includegraphics[height=6.5cm]{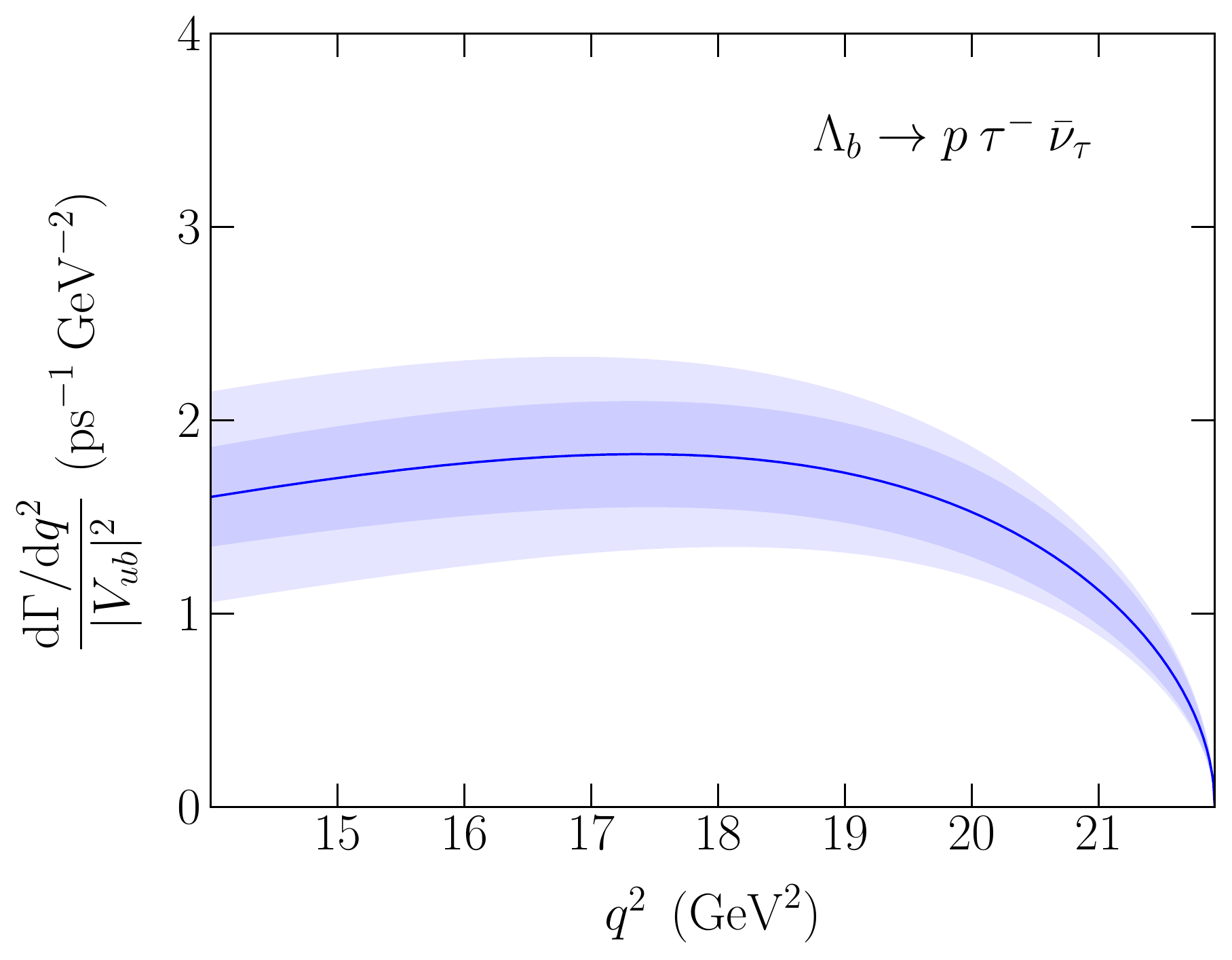}
\caption{\label{fig:dGamma}Our predictions for the differential decay rates of $\Lambda_b \to p\: \mu^- \bar{\nu}_\mu$ (left)
and $\Lambda_b \to p\: \tau^- \bar{\nu}_\tau$ (right), divided by $|V_{ub}|^2$. We only show the kinematic region where
we have lattice QCD results for the form factors $F_+$ and $F_-$. The inner error band originates from the statistical
plus systematic uncertainty in $F_\pm$. The outer error band additionally includes an estimate of the uncertainty caused
by the use of leading-order HQET for the $b$ quark. The plot for $\Lambda_b \to p\: e^- \bar{\nu}_e$ is indistinguishable
from $\Lambda_b \to p\: \mu^- \bar{\nu}_\mu$ and is therefore not shown.}
\end{figure}

In Fig.~\ref{fig:dGamma}, we show plots of $|V_{ub}|^{-2}\mathrm{d}\Gamma/\mathrm{d}q^2$ for the decays
$\Lambda_b \to p\: \mu^- \bar{\nu}_\mu$ and $\Lambda_b \to p\: \tau^- \bar{\nu}_\tau$ in the kinematic range where we
have lattice QCD results (in this range, the results for the electron final state look identical to the results for the
muon final state and are therefore not shown). The inner error bands in Fig.~\ref{fig:dGamma} originate from the total
uncertainty (statistical plus systematic) in the form factors $F_+$ and $F_-$. The use of leading-order HQET for the
$b$-quark introduces an additional systematic uncertainty in the differential decay rate, which is included in the outer
error band in Fig.~\ref{fig:dGamma}. At zero hadronic recoil, this uncertainty is expected to be of order
$\Lambda_{\rm QCD}/m_b$. At non-zero hadronic recoil, one further expects an uncertainty of order $|\mathbf{p'}|/m_b$,
because the proton momentum constitutes a new relevant scale. We add these two uncertainties in quadrature, and hence
estimate the systematic uncertainty in $|V_{ub}|^{-2}\mathrm{d}\Gamma/\mathrm{d}q^2$ that is caused by the use of
leading-order HQET to be 
\begin{equation}
 \sqrt{\frac{\Lambda_{\rm QCD}^2}{m_b^2}+\frac{|\mathbf{p'}|^2}{m_b^2}},
\end{equation}
where we take $\Lambda_{\rm QCD}=500$ MeV.

We also provide the following results for the integrated decay rate in the kinematic range of our lattice calculation,
$ 14\:{\rm GeV}^2 \leq q^2 \leq q^2_{\rm max}$ [where $q^2_{\rm max}=(m_{\Lambda_b}-m_N)^2$],
\begin{equation}
\label{eq:Gamma} \frac{1}{|V_{ub}|^2}\int_{14\:{\rm GeV}^2}^{q^2_{\rm max}}
\frac{\mathrm{d}\Gamma (\Lambda_b \to p\: \ell^- \bar{\nu}_\ell)}{\mathrm{d}q^2} \mathrm{d} q^2
 = \left\{ \begin{array}{ll} 15.3 \pm 2.4 \pm 3.4 \:\: \rm{ps}^{-1} & \hspace{1ex}{\rm for}\hspace{2ex}\ell=e, \\
                             15.3 \pm 2.4 \pm 3.4 \:\: \rm{ps}^{-1} & \hspace{1ex}{\rm for}\hspace{2ex}\ell=\mu, \\
                             12.5 \pm 1.9 \pm 2.7 \:\: \rm{ps}^{-1} & \hspace{1ex}{\rm for}\hspace{2ex}\ell=\tau.
            \end{array} \right.
\end{equation}
Here, the first uncertainty originates from the form factors, and the second uncertainty originates from the use of the static
approximation for the $b$-quark. With future experimental data, Eq.~(\ref{eq:Gamma}) can be used to determine $|V_{ub}|$.

\FloatBarrier
\section{\label{sec:conclusions}Discussion}
\FloatBarrier

We have obtained precise lattice QCD results for the $\Lambda_Q \to p$ form factors defined in the heavy-quark limit.
These results are valuable in their own right, as they can be compared to model-dependent studies performed in the same limit,
and eventually to future lattice QCD calculations at the physical $b$ quark mass.
For the $\Lambda_b \to p\: \ell^- \bar{\nu}_\ell$ differential decay rate, the static approximation introduces
a systematic uncertainty that is of order $\Lambda_{\rm QCD}/m_b\sim10\%$ at zero recoil and grows as the momentum
of the proton in the $\Lambda_b$ rest frame is increased. The total uncertainty for the integral of the differential decay
rate from $q^2=14\:{\rm GeV}^2$ to $q^2_{\rm max}=(m_{\Lambda_b}-m_N)^2$, which is the kinematic range where we have
lattice data, is about 30\%. Using future experimental data, this will allow a novel determination of the CKM matrix element
$|V_{ub}|$ with about 15\% theoretical uncertainty (the experimental uncertainty will also contribute to the overall extraction).
The theoretical uncertainty is already smaller than the difference between the values of $|V_{ub}|$ extracted from inclusive
and exclusive $B$ meson decays [Eqs.~(\ref{eq:Vubincl}) and (\ref{eq:Vubexcl})], and can be reduced further by performing
lattice QCD calculations of the full set of $\Lambda_b \to p$ form factors at the physical value of the $b$-quark mass.
In such calculations, the $b$ quark can be implemented using for example a Wilson-like action
\cite{ElKhadra:1996mp, Aoki:2001ra, Christ:2006us}, lattice nonrelativistic QCD \cite{Lepage:1992tx}, or higher-order
lattice HQET \cite{Blossier:2012qu}. Once the uncertainty from the static approximation is eliminated, other systematic
uncertainties need to be reduced. In the present calculation, the second-largest source of systematic uncertainty is the
one-loop matching of the lattice currents to the continuum current; ideally, in future calculations this can be replaced
by a nonperturbative method. We expect that after making these improvements, the theoretical uncertainty in the value of
$|V_{ub}|$ extracted from $\Lambda_b \to p\: \ell^- \bar{\nu}_\ell$ decays will be of order 5\%, and comparable to the
theoretical uncertainty for the analogous $\bar{B} \to \pi^+ \ell^- \bar{\nu}_\ell$ decays.

\FloatBarrier

\begin{acknowledgments}
We thank Ulrik Egede for a communication about the possibility of measuring $\Lambda_b \to p\: \mu^- \bar{\nu}_\mu$ at LHCb.
We thank the RBC and UKQCD collaborations for providing the gauge field configurations. This work has made use of the Chroma
software system for lattice QCD \cite{Edwards:2004sx}. WD and SM are supported by the U.S.~Department of Energy under
cooperative research agreement Contract Number DE-FG02-94ER40818. WD is also supported by U.S.~Department of Energy
Outstanding Junior Investigator Award DE-S{C0}0{0-17}84. CJDL is supported by Taiwanese NSC Grant Number 99-2112-M-009-004-MY3,
and MW is supported by STFC. Numerical calculations were performed using machines at NICS/XSEDE (supported by National
Science Foundation Grant Number OCI-1053575) and at NERSC (supported by U.S.~Department of Energy Grant Number DE-AC02-05CH11231).
\end{acknowledgments}


\begin{thebibliography}{10}

\bibitem{Kowalewski:2010zz} 
  R.~Kowalewski (BaBar Collaboration),
  PoS FPCP {\bf 2010}, 028 (2010).

\bibitem{Mannel:2010zz} 
  T.~Mannel,
  PoS FPCP {\bf 2010}, 029 (2010).

\bibitem{Antonelli:2009ws} 
  M.~Antonelli {\it et al.},
  Phys.\ Rept.\  {\bf 494}, 197 (2010)
  [arXiv:0907.5386].

\bibitem{PDG2012} 
  J.~Beringer {\it et al.}  (Particle Data Group Collaboration),
  Phys.\ Rev.\ D {\bf 86}, 010001 (2012).

\bibitem{Bailey:2008wp} 
  J.~A.~Bailey {\it et al.},
  Phys.\ Rev.\ D {\bf 79}, 054507 (2009)
  [arXiv:0811.3640].

\bibitem{Dalgic:2006dt}
  E.~Gulez, A.~Gray, M.~Wingate, C.~T.~H.~Davies, G.~P.~Lepage, and J.~Shigemitsu,
  Phys.\ Rev.\ D {\bf 73} (2006) 074502
   [Erratum-ibid.\ D {\bf 75} (2007) 119906]
  [arXiv:hep-lat/0601021].

\bibitem{Egede}
  U.~Egede, private communication.

\bibitem{Hussain:1992rb} 
  F.~Hussain, D.-S.~Liu, M.~Kramer, J.~G.~K\"orner, and S.~Tawfiq,
  Nucl.\ Phys.\ B {\bf 370}, 259 (1992).

\bibitem{Mannel:1990vg} 
  T.~Mannel, W.~Roberts, and Z.~Ryzak,
  Nucl.\ Phys.\ B {\bf 355}, 38 (1991).

\bibitem{Hussain:1990uu} 
  F.~Hussain, J.~G.~K\"orner, M.~Kramer, and G.~Thompson,
  Z.\ Phys.\ C {\bf 51}, 321 (1991).

\bibitem{Feldmann:2011xf} 
  T.~Feldmann and M.~W.~Y.~Yip,
  Phys.\ Rev.\ D {\bf 85}, 014035 (2012)
  [Erratum-ibid.\ D {\bf 86}, 079901 (2012)]
  [arXiv:1111.1844].

\bibitem{Mannel:2011xg} 
  T.~Mannel and Y.-M.~Wang,
  JHEP {\bf 1112}, 067 (2011)
  [arXiv:1111.1849].

\bibitem{Wang:2011uv} 
  W.~Wang,
  Phys.\ Lett.\ B {\bf 708}, 119 (2012)
  [arXiv:1112.0237].

\bibitem{Huang:1998rq}
  C.-S.~Huang, C.-F.~Qiao, and H.-G.~Yan,
  arXiv:hep-ph/9805452v3.

\bibitem{Carvalho:1999ia} 
  R.~S.~Marques de Carvalho, F.~S.~Navarra, M.~Nielsen, E.~Ferreira, and H.~G.~Dosch,
  Phys.\ Rev.\ D {\bf 60}, 034009 (1999)
  [arXiv:hep-ph/9903326].

\bibitem{Huang:2004vf} 
  M.-Q.~Huang and D.-W.~Wang,
  Phys.\ Rev.\ D {\bf 69}, 094003 (2004)
  [arXiv:hep-ph/0401094].

\bibitem{Wang:2009hra} 
  Y.-M.~Wang, Y.-L.~Shen, and C.-D.~L\"u,
  Phys.\ Rev.\ D {\bf 80}, 074012 (2009)
  [arXiv:0907.4008].

\bibitem{Azizi:2009wn} 
  K.~Azizi, M.~Bayar, Y.~Sarac, and H.~Sundu,
  Phys.\ Rev.\ D {\bf 80}, 096007 (2009)
  [arXiv:0908.1758].

\bibitem{Khodjamirian:2011jp} 
  A.~Khodjamirian, C.~Klein, T.~Mannel, and Y.-M.~Wang,
  JHEP {\bf 1109}, 106 (2011)
  [arXiv:1108.2971].

\bibitem{Bahr:2012vt} 
  F.~Bahr, F.~Bernardoni, A.~Ramos, H.~Simma, R.~Sommer, and J.~Bulava,
  PoS LATTICE {\bf 2012}, 110 (2012)
  [arXiv:1210.3478].

\bibitem{Bouchard:2012tb} 
  C.~M.~Bouchard, G.~P.~Lepage, C.~J.~Monahan, H.~Na and J.~Shigemitsu,
  PoS LATTICE {\bf 2012}, 118 (2012)
  [arXiv:1210.6992].

\bibitem{Kawanai:2012id} 
  T.~Kawanai, R.~S.~Van de Water, and O.~Witzel,
  PoS LATTICE {\bf 2012}, 109 (2012)
  [arXiv:1211.0956].

\bibitem{Detmold:2012vy} 
  W.~Detmold, C.-J.~D.~Lin, S.~Meinel, and M.~Wingate,
  Phys.\  Rev.\  D 87, {\bf 074502} (2013)
  [arXiv:1212.4827].

\bibitem{Kaplan:1992bt}
  D.~B.~Kaplan,
  Phys.\ Lett.\  B {\bf 288}, 342 (1992)
  [arXiv:hep-lat/9206013].

\bibitem{Shamir:1993zy}
  Y.~Shamir,
  Nucl.\ Phys.\  B {\bf 406}, 90 (1993)
  [arXiv:hep-lat/9303005].

\bibitem{Furman:1994ky}
  V.~Furman and Y.~Shamir,
  Nucl.\ Phys.\  B {\bf 439}, 54 (1995)
  [arXiv:hep-lat/9405004].

\bibitem{Aoki:2010dy}
  Y.~Aoki {\it et al.}  (RBC/UKQCD Collaboration),
  Phys.\ Rev.\  D {\bf 83}, 074508 (2011)
  [arXiv:1011.0892].

\bibitem{Iwasaki:1983ck}
  Y.~Iwasaki,
  Report No. UTHEP-118 (1983).

\bibitem{Iwasaki:1984cj}
  Y.~Iwasaki and T.~Yoshie,
  Phys.\ Lett.\  B {\bf 143}, 449 (1984).

\bibitem{Eichten:1989kb}
  E.~Eichten and B.~R.~Hill,
  Phys.\ Lett.\  B {\bf 240}, 193 (1990).

\bibitem{DellaMorte:2005yc} 
  M.~Della Morte, A.~Shindler, and R.~Sommer,
  JHEP {\bf 0508}, 051 (2005)
  [arXiv:hep-lat/0506008].

\bibitem{Ishikawa:2011dd} 
  T.~Ishikawa, Y.~Aoki, J.~M.~Flynn, T.~Izubuchi, and O.~Loktik,
  JHEP {\bf 1105}, 040 (2011)
  [arXiv:1101.1072].

\bibitem{Jenkins:1991ne}
  E.~E.~Jenkins and A.~V.~Manohar,
  UCSD-PTH-91-30,
  Proceedings of the workshop on Effective Field Theories of the Standard Model,
  Dobog\'ok\H{o}, Hungary, August 1991.

\bibitem{Jenkins:1990jv}
  E.~E.~Jenkins and A.~V.~Manohar,
  Phys.\ Lett.\  B {\bf 255}, 558 (1991).

\bibitem{Yan:1992gz}
  T.-M.~Yan, H.-Y.~Cheng, C.-Y.~Cheung, G.-L.~Lin, Y.~C.~Lin, and H.-L.~Yu,
  Phys.\ Rev.\  D {\bf 46}, 1148 (1992)
  [Erratum-ibid.\  D {\bf 55}, 5851 (1997)].

\bibitem{Cho:1992cf}
  P.~L.~Cho,
  Nucl.\ Phys.\  B {\bf 396}, 183 (1993)
  [Erratum-ibid.\  B {\bf 421}, 683 (1994)]
  [arXiv:hep-ph/9208244].

\bibitem{Cabibbo:1963yz} 
  N.~Cabibbo,
  Phys.\ Rev.\ Lett.\  {\bf 10}, 531 (1963).

\bibitem{Kobayashi:1973fv} 
  M.~Kobayashi and T.~Maskawa,
  Prog.\ Theor.\ Phys.\  {\bf 49}, 652 (1973).

\bibitem{Wilson:1969zs} 
  K.~G.~Wilson,
  Phys.\ Rev.\  {\bf 179}, 1499 (1969).

\bibitem{Eichten:1989zv} 
  E.~Eichten and B.~R.~Hill,
  Phys.\ Lett.\ B {\bf 234}, 511 (1990).

\bibitem{ElKhadra:1996mp} 
  A.~X.~El-Khadra, A.~S.~Kronfeld, and P.~B.~Mackenzie,
  Phys.\ Rev.\ D {\bf 55}, 3933 (1997)
  [arXiv:hep-lat/9604004].

\bibitem{Aoki:2001ra} 
  S.~Aoki, Y.~Kuramashi, and S.-i.~Tominaga,
  Prog.\ Theor.\ Phys.\  {\bf 109}, 383 (2003)
  [arXiv:hep-lat/0107009].

\bibitem{Christ:2006us} 
  N.~H.~Christ, M.~Li, and H.-W.~Lin,
  Phys.\ Rev.\ D {\bf 76}, 074505 (2007)
  [arXiv:hep-lat/0608006].

\bibitem{Lepage:1992tx}
  G.~P.~Lepage, L.~Magnea, C.~Nakhleh, U.~Magnea, and K.~Hornbostel,
  Phys.\ Rev.\  D {\bf 46}, 4052 (1992)
  [arXiv:hep-lat/9205007].

\bibitem{Blossier:2012qu} 
  B.~Blossier {\it et al.}  (ALPHA Collaboration),
  JHEP {\bf 1209}, 132 (2012)
  [arXiv:1203.6516].

\bibitem{Edwards:2004sx}
  R.~G.~Edwards and B.~Jo\'o,
  Nucl.\ Phys.\ Proc.\ Suppl.\  {\bf 140}, 832 (2005)
  [arXiv:hep-lat/0409003].

\end{thebibliography}
\end{document}